\documentclass[aps,prc,twocolumn,nofootinbib,superscriptaddress,showpacs,floatfix,preprintnumbers]{revtex4-2}

\usepackage[dvipsnames]{xcolor}     
\definecolor{lcolor}{rgb}{0.5,0,0}
\definecolor{citcolor}{rgb}{0,0,1}
\usepackage[breaklinks,colorlinks,urlcolor=blue,citecolor=citcolor,linkcolor=lcolor,linktoc=all]{hyperref}
\usepackage{color}
\usepackage{graphicx}	
\graphicspath{{./figures/}}
\usepackage[utf8]{inputenc}
\usepackage{amsmath} \usepackage{amssymb}
\usepackage[dvipsnames]{xcolor}      
\usepackage[capitalise]{cleveref}
\usepackage{bm,bbm,bbold}
\usepackage{ulem}
\usepackage{placeins}

\allowdisplaybreaks

\makeatletter
\g@addto@macro\bfseries{\boldmath}
\makeatother

\usepackage{tikz}
\usepackage[customcolors]{hf-tikz}
\usepackage{mciteplus}

\usetikzlibrary{arrows,cd,shapes,decorations.pathmorphing,decorations.markings,shadings}
\tikzset{
  big arrow/.style={
    decoration={markings,mark=at position 1 with {\arrow[scale=4,#1]{>}}},
    postaction={decorate},
    shorten >=0.4pt},
  big arrow/.default=blue}

\newcommand{\be}{\begin{equation}}
\newcommand{\ee}{\end{equation}}
\newcommand{\bea}{\begin{eqnarray}}
\newcommand{\eea}{\end{eqnarray}}

\begin{document}

\title{The Interplay Between Electromagnetic Fields and Baryon Stopping in a Hydrodynamic Model for Charged Flow}
\author{Tuna Demircik}
\email{t.demircik@uu.nl}
\affiliation{ Institute for Theoretical Physics, Utrecht University, Leuvenlaan 4, 3584 CE Utrecht, The Netherlands}
\author{Dmitri E. Kharzeev}
\email{dmitri.kharzeev@uconn.edu}
\affiliation{Department of Physics, University of Connecticut, Storrs, Connecticut 06269, USA}
\author{Krishna Rajagopal}
\email{krishna@mit.edu}
\affiliation{MIT Center for Theoretical Physics — a Leinweber Institute, Massachusetts Institute of Technology, Cambridge MA 02139, USA}
\author{Raimond Snellings}
\email{r.snellings@uu.nl}
\affiliation{ Institute for Theoretical Physics, Utrecht University, Leuvenlaan 4, 3584 CE Utrecht, The Netherlands}
\affiliation{Nikhef, Science Park 105, 1098 XG Amsterdam, The Netherlands}

\preprint{MIT-CTP/6108}

\begin{abstract}
Charge-dependent directed flow provides a sensitive probe of early electromagnetic fields and baryon stopping in relativistic heavy-ion collisions. Recent STAR measurements show a centrality-dependent sign change in the directed flow splitting (the difference between the directed flow of protons and antiprotons), indicating that electromagnetic effects alone are not sufficient to describe this observable and that the 
baryon stopping --- in particular the component of the stopped proton distribution that is odd in rapidity and odd under reflection in the impact 
parameter direction --- must also be included. We develop a semi-analytic hydrodynamic framework that combines spectator-induced electromagnetic fields with a Glauber-based description of baryon stopping, built upon an analytic solution for the background hydrodynamic flow due to Gubser together with the simplifying assumption of a constant electrical conductivity. For Au+Au collisions at $\sqrt{s_{NN}}=200$ GeV, we find that baryon stopping gives a positive contribution to the directed flow splitting that decreases for more peripheral collisions, while electromagnetic fields give a negative contribution that is larger for more  peripheral collisions. The competition between these two effects naturally reproduces the observed sign change in Au+Au collisions as a function of
centrality,
describes the observed rapidity dependence in the $50$--$80\%$ centrality interval, and reproduces trends seen in U+U collisions. Although the simplifying assumptions that we have made regarding
the analytic background and constant conductivity
limit our ability to make quantitative comparisons, our model provides a transparent explanation of how transported baryon number and spectator-induced electromagnetic fields jointly shape charge-dependent directed flow.
\end{abstract}

\maketitle

\section{Introduction}

Ultra-relativistic heavy-ion collisions with a nonzero impact parameter generate extremely large electromagnetic fields, primarily sourced by the fast-moving spectator protons. Estimates based on the Biot-Savart law indicate that these fields can reach $10^{18}$--$10^{19}$ Gauss during the earliest stages of the collision~\cite{Kharzeev:2007jp,Skokov:2009qp,Tuchin:2013ie}. Such fields are of particular interest because they may induce anomalous transport phenomena in the presence of chirality imbalance, including the chiral magnetic effect, namely an electric current along the magnetic field~\cite{Kharzeev:2007jp,Fukushima:2008xe}, 
and the chiral magnetic wave, a collective mode generated by the coupling between electric and chiral charge density waves~\cite{Kharzeev:2010gd,Burnier:2011bf}. However, the charge-dependent observables~\cite{Voloshin:2004vk} proposed to search for these effects can also receive background contributions from conventional collective dynamics~\cite{Voloshin:2010ut,Schlichting:2010qia,Bzdak:2012ia,STAR:2021mii,Kharzeev:2022hqz}.
It is therefore imperative to calibrate the strength and lifetime of the early magnetic field independently~\cite{Gursoy:2014aka,Gursoy:2018yai,Dubla:2020bdz}.

To this end, directed flow ($v_1$) has been identified as the relevant observable~\cite{Gursoy:2014aka,Gursoy:2018yai}. It is the first harmonic coefficient in the Fourier expansion of the final-state-hadron azimuthal distribution with respect to the reaction plane, and quantifies the collective sideward motion of produced particles~\cite{Voloshin:1994mz,Poskanzer:1998yz,Bilandzic:2010jr}. In symmetric collisions, $v_1(y)$ is odd in rapidity and is therefore commonly characterized near mid-rapidity by its slope $dv_1/dy$. Since directed flow is generated by the early sideward dynamics of the fireball, 
differences between the directed flow of hadrons that have opposite electric charge but are otherwise identical are a natural consequence of electromagnetic forces.
Following Refs.~\cite{Gursoy:2014aka,Gursoy:2018yai},
we shall therefore focus on the charge-dependent directed flow observables
$\Delta v_1\equiv v_1(h^+)-v_1(h^-)$ and
$\Delta(dv_1/dy)\equiv dv_1(h^+)/dy-dv_1(h^-)/dy$.
By looking at the splitting between the directed flow of  hadrons that are identical 
other than by virtue of having opposite charge we also minimize the sensitivity of the observables we investigate to details of the background hydrodynamic flow. 

\begin{figure}[t]
\centering
\includegraphics[width=1.\columnwidth]{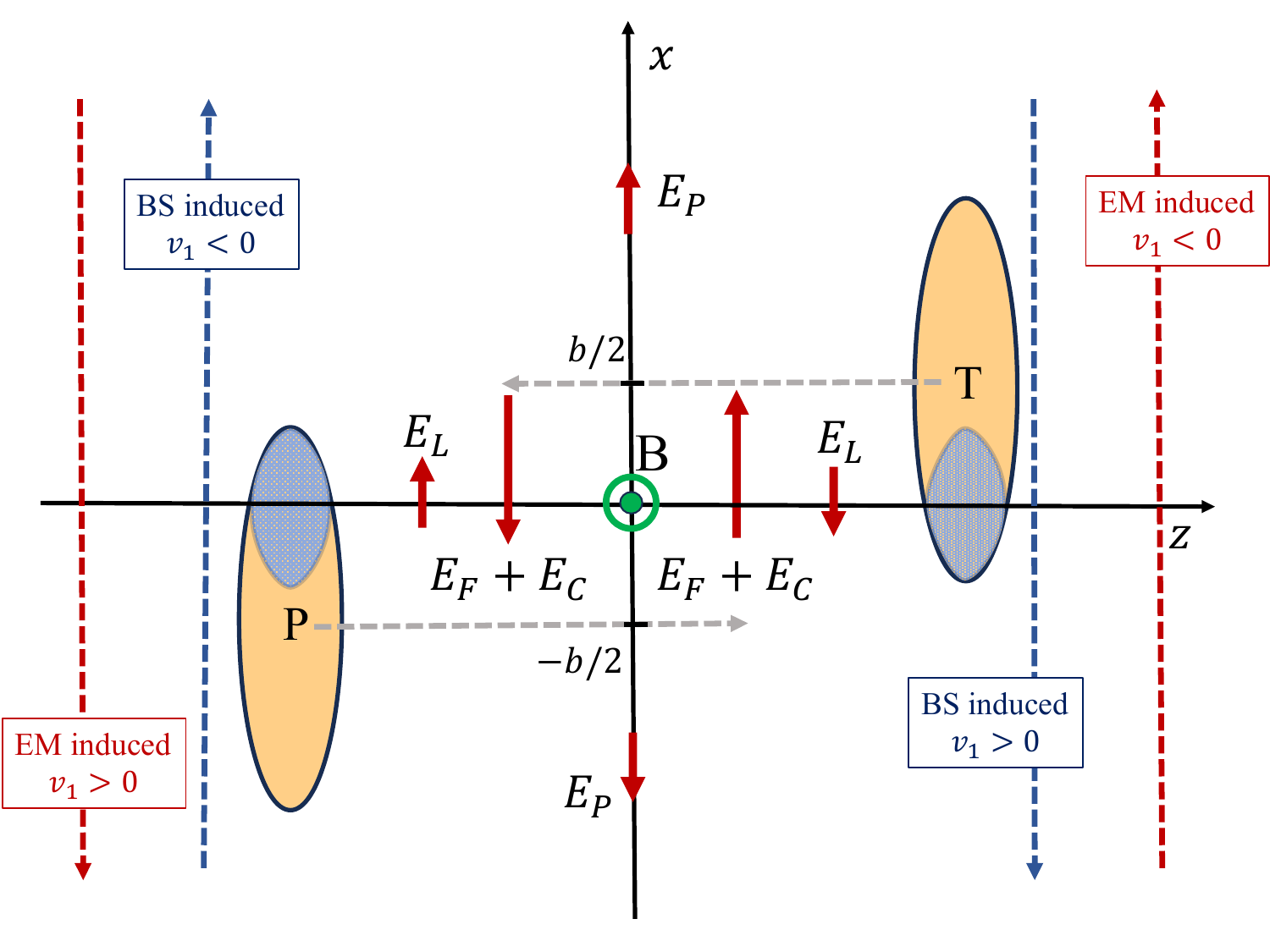}
\caption{Schematic illustration of the collision geometry and the electromagnetic fields relevant for understanding the four mechanisms described in the text that result in charge-dependent directed flow. The beam axis is along the $z$ direction and the impact parameter is along the $x$ direction. 
With the convention used here, the ``projectile'' nucleus P moves toward positive $z$ and is centered at $x=-b/2$, while the ``target'' nucleus T moves toward negative $z$ and is centered at $x=+b/2$. Each is a Lorentz-contracted
pancake. The blue almond-shaped regions of the incident nuclei represent the participants that are about to collide;
the orange regions represent the spectators that will continue onwards, meaning that just after the collision the orange spectators from T/P will be found at positive/negative $x$ and negative/positive $z$.
The passing spectators create a magnetic field $B_y$ pointing out of the $(x,z)$-plane, depicted as a green vector.
The red vectors indicate the 
directions of the induced electric fields and resulting currents 
originating from the four electromagnetic mechanisms described in the text: Faraday current, Coulomb current, Lorentz current, and Plasma current. 
The red dashed arrows show the direction of the corresponding rapidity-odd directed flow pattern, assuming that the Faraday and Coulomb contributions dominate over the Lorentz contribution.
In the collision zone, 
because the edge of one nucleus collides with a region closer to the middle of the other nucleus and vice versa
the nonzero density
of stopped baryons (with a net positive charge from the protons in the incident nuclei) 
accumulates such that its distribution has a component that is odd under both $z\rightarrow -z$ (odd in rapidity)
and under $x\rightarrow -x$ (which we shall refer to as reflection-odd).
The resulting rapidity-odd and reflection-odd
positive charge
is then carried outward by the radial expansion of the droplet of QGP. 
The blue dashed arrows show 
the direction of the corresponding 
rapidity-odd directed flow pattern.}
\label{coll_geo}
\end{figure}
%%%%%%%%%%%%%%%%%%%%%%%

Theoretical studies have provided the baseline expectation for such electromagnetically induced directed flow. In the pioneering work in this direction~\cite{Gursoy:2014aka}, the  background flow of the hydrodynamic droplet of QGP produced in a heavy ion collision was modeled qualitatively using an analytic solution first found by Gubser~\cite{Gubser:2010ze}, 
which made it possible to obtain a simple, perturbative, and largely analytic treatment of the electromagnetic fields, the resulting drift of charged particles, and the consequent charge-dependent directed flow.  In this framework, the charge-dependent directed flow receives contributions from four ordinary electromagnetic mechanisms, which are illustrated schematically  in Fig.~\ref{coll_geo}. These mechanisms are as follows: (1) \textit{Faraday current $(E_F)$:} The magnetic field perpendicular to the reaction plane (in the $+y$ direction in Fig.~\ref{coll_geo}) produced at the earliest moments of the collision by the passing spectators subsequently decreases with time, inducing 
an electric current in the positive/negative $x$ direction at positive/negative $z$
through Faraday's law; 
(2) \textit{Couloumb current $(E_C)$:} Just after the collision, the spectator protons are at positive/negative $x$ for
negative/positive $z$; these spectators generated a Coulomb electric field that drives a current in the same direction 
as the Faraday current;
(3) \textit{Lorentz current $(E_L)$:} The longitudinal expansion of the positively charged QGP produced in the collision
constitutes a current in the positive/negative $z$ direction at positive/negative $z$; in the presence of the magnetic field, the Lorentz force then drives a current, often called a Hall current, in the opposite direction as the Faraday current;
(4) \textit{Plasma current $(E_P)$,} which we mention for completeness, although it does not contribute to the charge-dependent {\it directed} flow: The net positive charge of the plasma produces an outward electric field, giving rise to charge-dependent radial and elliptical flow. 

For the rapidity-odd directed-flow splitting, the Faraday and Coulomb contributions act against the Lorentz contribution, and existing calculations typically find the former to dominate, leading to a \textit{negative} midrapidity slope of $\Delta v_1$~\cite{Gursoy:2014aka,Gursoy:2018yai}. The same perturbative strategy introduced in Ref.~\cite{Gursoy:2014aka} was later implemented on top of more realistic viscous hydrodynamic simulations~\cite{Gursoy:2018yai}, confirming the qualitative origin of the charge-odd, rapidity-odd $v_1$ signal while improving the description of the underlying bulk evolution. The calculation of the charge-dependent directed flow resulting from electromagnetic forces has been further advanced in Ref.~\cite{Benoit:2025amn}.

Experimentally, charge-dependent directed flow measurements for identified hadrons have been reported over a broad range of RHIC beam energies. In collisions at the top RHIC energy $\sqrt{s_{NN}}=200$ GeV, the available pion data from STAR are 
are consistent with
the negative $\Delta v_1$ slope expected from electromagnetic-field calculations, 
but they are also consistent with zero~\cite{STAR:2023jdd}.
For protons, the situation is very interestingly different.  As we noted above, 
calculations of the effects of electromagnetic fields predict
a negative slope in rapidity for
$\Delta v_1(p)\equiv v_1(p)-v_1(\bar p)$. In contrast, STAR observes a positive 
$d \Delta v_1(p)/dy$ in
collisions with centrality $\lesssim 50\%$ and a negative 
$d \Delta v_1(p)/dy$ 
only in more peripheral collisions~\cite{STAR:2023jdd}.
As noted by the STAR collaboration~\cite{STAR:2023jdd},
this highlights the fact that there is an additional contribution to $d \Delta v_1/dy$ specifically for protons.
Some protons from the incident nuclei 
are stopped during a heavy ion collision and end up within the droplet of QGP produced in the collision.
In collisions with nonzero impact parameter as diagrammed in Fig.~\ref{coll_geo}, the 
deposition of stopped protons
is odd under change of sign of  rapidity 
and odd under reflection in the impact parameter axis $x$.
At positive $x$  in Fig.~\ref{coll_geo},
the upper-edge of the right-going nucleus P collides with the interior of the left-going nucleus T, which means that the stopped protons will tend to have negative rapidity.  
At negative $x$, the stopped protons will tend to have positive rapidity.  The subsequent outward radial flow of the expanding droplet of QGP will then turn this rapidity-asymmetric and $x$-asymmetric density of stopped protons into a rapidity-asymmetric directed flow of protons, with $v_1>0$ at positive rapidity and $v_1<0$ at negative 
rapidity. Since there is no analogous effect for antiprotons, the result is a positive contribution to the rapidity slope of the charge-dependent directed flow $d \Delta v_1(p)/dy$.
This effect, which 
was first discussed in Ref.~\cite{Snellings:1999bt},
has been seen in 
UrQMD simulations~\cite{Guo:2012qi}, AMPT simulations~\cite{Nayak:2019vtn}, and in hydrodynamic simulations with an inhomogeneous baryon density~\cite{Bozek:2022svy,Parida:2025ddt}, all of which confirm that the stopped protons together with radial flow result in a
positive contribution to the directed-flow slope for protons-antiprotons. 

In Ref.~\cite{STAR:2023jdd}, the STAR collaboration suggested that the effects of electromagnetic fields in combination with the contribution from stopped protons
could explain the change in sign of $d \Delta v_1/dy$ with centrality seen in their experimental measurements --- with the contribution from stopped protons more (less) important than that from electromagnetic forces in central (peripheral) collisions where the magnetic field produced by the spectators is smallest (largest).
Our goal in this paper is to construct a model that incorporates both electromagnetic effects and baryon stopping within a single framework, and to confront the results that we obtain with the STAR data.

In Refs.~\cite{STAR:2023jdd,Taseer:2024sho}, STAR has reported many more measurements of charge-dependent directed flow.  They reported
charge-dependent measurements of $dv_1/dy$ for $\pi^\pm$, $K^\pm$, and $p(\bar p)$ in Au+Au~\cite{STAR:2023jdd}, 
Ru+Ru and Zr+Zr~\cite{STAR:2023jdd}, and U+U~\cite{Taseer:2024sho} collisions at $\sqrt{s_{NN}}=200$~GeV, as well as in Au+Au collisions at $\sqrt{s_{NN}}=27$~GeV~\cite{STAR:2023jdd}.  In all of these collision systems,
$d\Delta v_1/dy$ for both $p-\bar p$
and $K^+-K^-$ show the same nontrivial centrality dependence,
with the rapidity-slope of the charge-dependent directed flow changing from positive for central-midcentral collisions to negative for midcentral-peripheral collisions.
In all these cases, the magnitude
of $d\Delta v_1/dy$ for $\pi^+-\pi^-$ is smaller, typically consistent with zero within error bars.

We furthermore note that
STAR has also measured the rapidity dependence of $v_1$ and its mid-rapidity slope $dv_1/dy$ for $p$, $\bar{p}$, $\pi^+$ and $\pi^-$ in Au+Au collisions from the Beam Energy Scan (BES) for collision energies $\sqrt{s_{NN}}$ between 7.7 and $200$ GeV~\cite{STAR:2014clz,STAR:2017okv,Taseer:2024sho}. The proton and antiproton $dv_1/dy$ differ substantially at low collision energies, but this difference decreases as the beam energy increases. The proton $dv_1/dy$ changes sign from positive to negative between $\sqrt{s_{NN}}=7.7$ and $11.5$ GeV, reaches a minimum between $11.5$ and $19.6$ GeV, and remains small and negative up to top RHIC collision energies. In contrast, the $dv_1/dy$ of $\pi^+$ and $\pi^-$ are negative and similar in magnitude in high energy collisions, 
with only small charge-dependent differences at the lowest BES energies.
The model for baryon stopping that we shall employ as well as the boost-invariant hydrodynamics in our modeling are not well-suited to analyzing these lower collision energies, which we leave to future work. 

At the LHC, ALICE has measured a positive charge-dependent slope for inclusive charged hadrons in Pb+Pb collisions at $\sqrt{s_{NN}}=5.02$ TeV with a significance of about $2.6\sigma$, while model calculations for charged pions predict a slope of similar magnitude but opposite sign~\cite{ALICE:2019sgg}. This comparison remains inconclusive because the measurement is not yet separated by hadron species; future identified-particle measurements will therefore be crucial for clarifying the origin of the observed charge-dependent directed flow.

Taken together, these experimental results, particularly the change
in the sign of the rapidity-slope of the charge-dependent
directed flow from positive for central collisions to negative for peripheral collisions seen for both protons and kaons in many collision systems,
suggest that both stopped protons and
electromagnetic forces are important,
with their contributions changing
in importance as a function of centrality.
Baryon stopping provides a natural baseline in more central events, while the Faraday and Coulomb contributions are expected to become much more important toward peripheral collisions where magnetic fields are larger. 
This makes it imperative to develop
a unified framework in which both the electromagnetic response and the 
contribution from baryon stopping are treated consistently.
In this study, we aim to address this gap between theory and experiment by constructing a model that incorporates both electromagnetic effects and baryon stopping within a single framework. 

As in the original work of Ref.~\cite{Gursoy:2014aka} we build our treatment upon
the analytic Gubser solution to relativistic viscous hydrodynamics~\cite{Gubser:2010ze} and also rely upon additional simplifying assumptions including treating the electrical conductivity of QGP as if it were constant even though it is expected to increase at least linearly with $T$.
The Gubser solution has well-known limitations: (i) the longitudinal expansion is boost-invariant, which is a good approximation over a range of mid-rapidities in high energy collisions but cannot describe lower energy collisions in the RHIC Beam Energy Scan; and (ii) 
the Gubser solution is azimuthally symmetric, which means that it cannot describe noncentral collisions 
in detail.
Nevertheless, because the Gubser solution is analytic, it enables a semi-analytic treatment of both baryon stopping and electromagnetic fields and forces. 
This makes it useful for incorporating and isolating the electromagnetic and baryon-stopping contributions consistently, while keeping their interplay physically transparent, as is our goal in this work.  
Future quantitative predictions for experimental data will need to be built upon state-of-the art hydrodynamic simulations of the droplets of QGP produced in non-central heavy
ion collisions, as was done for
the contributions from electromagnetic 
forces in Refs.~\cite{Gursoy:2018yai,Parida:2025ddt}, and will need to take account of the temperature-dependence of the electrical conductivity of QGP.

Our model extends the analysis of Refs.~\cite{Gursoy:2014aka,Gursoy:2018yai} in two main directions. First, it incorporates baryon stopping by introducing a nontrivial baryon chemical potential associated with a stopping profile. Second, while Ref.~\cite{Gursoy:2014aka} focused on collisions in the $20$--$30\%$ centrality class, we extend the analysis to the full centrality range covered by the existing data. We do this by varying the impact parameter in the electromagnetic-field and baryon-stopping calculations, where it enters explicitly. As the Gubser background is azimuthally symmetric,
we cannot literally introduce an
impact parameter, but we mimic
the impact-parameter dependence of the
fireball in a crude way by employing
smaller diameter Gubser solutions with increasing impact parameter.
This captures part of the expected centrality dependence, but 
a model built upon the Gubser solution
cannot yield a quantitative description.
These limitations, together with the strategies that we use to compensate for them and the other approximations made in the model, will be discussed 
at appropriate points 
throughout the rest of this work.

Our focus is on the directed flow of protons and antiprotons in Au+Au collisions at $\sqrt{s_{NN}}=200$ GeV at RHIC. We find that, despite its limitations, the model reproduces the main centrality-dependent trends observed in Refs.~\cite{STAR:2023jdd,Taseer:2024sho} rather well, as shown in Figs.~\ref{res03}, \ref{res002}, and \ref{res001}.  
We re-emphasize, however, that the outputs of our model cannot be viewed as quantiative predictions for experimental data,
given the limitations of the Gubser solution and the additional simplifying assumptions adopted here. 
With this caveat in mind, our semi-analytic model should be viewed as a simple and transparent tool for studying the interplay between electromagnetic effects and baryon stopping. 
In particular, it provides a  satisfying, qualitative, explanation of the apparent mismatch between previous electromagnetic-field predictions and experimental measurements, including the centrality-dependent sign change of the rapidity-slope of the charge-dependent directed flow.

The rest of this paper is organized as follows. In Sec.~II, we introduce the hydrodynamic background based on Gubser flow that we shall employ
and in Sec.~III we specify the Cooper--Frye freezeout prescription used throughout the calculation. In Sec.~IV, we describe how baryon stopping is implemented through a spacetime-dependent baryon chemical potential constructed from a Glauber-based stopped-baryon profile. In Sec.~V, we summarize the construction of the electromagnetic field generated by the charged spectators in a conducting medium. In Sec.~VI, we explain how the electromagnetic drift and the baryon-stopping reweighting are combined to compute the directed flow of protons and antiprotons. The main results are presented in Sec.~VII, followed by a discussion of the physical interpretation, limitations, and outlook in Sec.~\ref{sec:Conclusion}.

\section{Hydrodynamic background: Gubser flow}
\label{sec:HydroBackground}

As the hydrodynamic background, we use the analytic solution found by Gubser for a conformal fluid~\cite{Gubser:2010ze}. This solution describes a finite-size droplet of hot, conformal, strongly coupled, low viscosity,  gauge theory plasma undergoing boost-invariant expansion along the beam direction together with nontrivial transverse expansion. It is constructed by imposing boost invariance in the longitudinal direction and rotational symmetry in the transverse plane, together with  additional conformal symmetries and reflection symmetry under $z\leftrightarrow-z$. In other words, Gubser solution preserves  $SO(1,1)\times SO(3)\times Z_2$. As a result, the flow is azimuthally symmetric and invariant under reflection along the beam axis. Gubser flow may be viewed as a simplified, highly symmetric, model for the expanding, cooling, droplet of quark-gluon plasma created in a heavy-ion collision.

In Milne coordinates \(x^\mu=(\tau,x_\perp,\phi,\eta)\), the nonvanishing components of the fluid four-velocity  are
\begin{align}
\label{eq:u-Milne}
u^\tau &= \frac{1+q^2\tau^2+q^2x_\perp^2}{2q\tau\sqrt{1+g^2}}\ , \qquad
u^\perp = \frac{q x_\perp}{\sqrt{1+g^2}}\ ,
\end{align}
with
\[
g \equiv \frac{1+q^2x_\perp^2-q^2\tau^2}{2q\tau}.
\]
An important feature of the solution is that the flow profile is controlled by a single parameter \(q\) which has dimensions of inverse-length, meaning that it sets the initial transverse size of the droplet of fluid described by the Gubser solution. 

The local temperature profile contains two pieces:
\begin{equation}
T(\tau,x_\perp)=T_{\rm ideal}(\tau,x_\perp)+\delta T_{\rm visc}(\tau,x_\perp),
\label{eq:TempProfile}
\end{equation}
with
\begin{align}
T_{\rm ideal}(\tau,x_\perp)
&=\frac{\hat T_0}{\tau\, f_*^{1/4}(1+g^2)^{1/3}}, 
\label{eq:Tideal}
\\
\label{eq:Tvisc}
\delta T_{\rm visc}(\tau,x_\perp)
&=\frac{1}{\tau\, f_*^{1/4}}
\frac{H_0\, g}{\sqrt{1+g^2}} \\
&\;\;\;\;\times
\left(
1-(1+g^2)^{1/6}\,
{}_2F_1\!\left(\frac12,\frac16;\frac32;-g^2\right)
\right)\ ,. \nonumber
\end{align}
where $_2F_1$ is the ordinary hypergeometric function. The first term
in Eq.~\eqref{eq:TempProfile}, proportional to \(\hat T_0\), is the ideal-fluid contribution. 
For a conformal fluid, the local energy density $\varepsilon$ is related to the temperature $T$ by
\begin{equation}
\label{eq:EnergyDensity}
\varepsilon(\tau,x_\perp)=f_*\,T^4(\tau,x_\perp),
\end{equation}
where \(f_*\) is a constant that fixes the equation of state.
We take $f_*=11$, which is reasonable for QCD thermodynamics at temperatures around 250~MeV~\cite{Borsanyi:2013bia,HotQCD:2014kol}.
Specifying $u^\mu$ as in Eq.~\eqref{eq:u-Milne} and $T$ as in Eqs.~\eqref{eq:TempProfile}-\eqref{eq:Tvisc} completes the specification of our hydrodynamic background.
With the energy density specified via the choice of $f_*$ in Eq.~\eqref{eq:EnergyDensity}, the pressure and entropy density and 
all other thermodynamic quantities follow.
The second term in Eq.~\eqref{eq:TempProfile} 
is the dissipative correction, controlled by the parameter \(H_0\), which represents the effect of shear viscosity on the temperature profile via 
$\eta= H_0 \varepsilon^{3/4}$. 
Thus, \(\hat T_0\) fixes the overall temperature scale, while \(H_0\) determines the size of 
viscous effects~\cite{Gubser:2010ze}.
Following Gubser~\cite{Gubser:2010ze}, we fix  $H_0=0.33$ which corresponds to $\eta/s\simeq 0.134$, 
which lies within the range of values extracted via Bayesian uncertainty quantification from comparison between sophisticated hydrodynamic calculations and experimental measurements of many highly differential anisotropic flow 
observables~\cite{Nijs:2020ors,Nijs:2020ors,JETSCAPE:2020mzn,Parkkila:2021tqq,Parkkila:2021yha}.
We shall describe how we fix the parameters $q$ and $\hat T_0$ further below. 

Although Gubser flow is too symmetric to describe a realistic non-central collision in detail, it is still very useful as an analytically controlled background. In particular, with a suitable choice of parameters, it gives a reasonable qualitative description of hadron spectra while keeping the calculation simple~\cite{Gursoy:2014aka}. For our purposes, this is sufficient, since we are not trying to model the full geometry-driven anisotropic flow of a noncentral collision. Instead, we are interested in the much smaller rapidity-odd directed flow signal driven by electromagnetic forces and the small number of stopped protons. In this setting, Gubser flow provides a convenient and physically transparent starting point. We shall introduce the electromagnetic fields as perturbations that modify the Gubser flow (Sec.~\ref{EMcomp}), and shall model the effects of stopped baryons via a baryon chemical potential that reweights the Cooper-Frye freezeout surface (Sec.~\ref{BScomp}). 

\subsection{Fixing the Gubser parameters $q$ and $\hat T_0$}

\begin{figure}[t]
\centering
\includegraphics[width=1.\columnwidth]{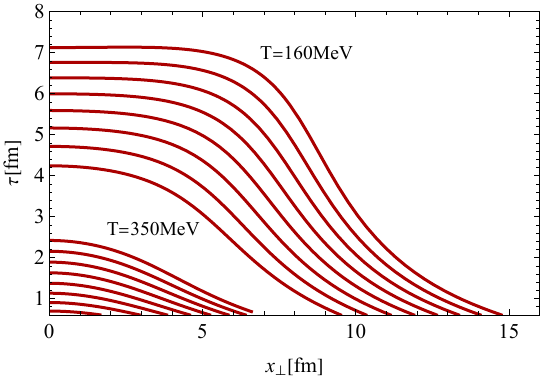}
\caption{Isotherms at $T=350~\mathrm{MeV}$
and $T=160$~MeV 
for the Gubser solutions that we use to model 
Au+Au collisions with 
$\sqrt{s_{\rm NN}}=200$~GeV
and (top to bottom) 5\%, 10\%, 20\%, 30\%, 40\%, 50\%, 60\% and 70\% 
centralities. 
Note that we shall take the freezeout temperature to be $T_f=160$~MeV, meaning that the 160 MeV isotherms are the 
freezeout surfaces that we shall employ in the next Section.
}
\label{res01}
\end{figure}

As explained above, the Gubser solution does not contain an explicit impact-parameter dependence. Our strategy for mimicking centrality dependence is therefore to assign a different effective Gubser background, with a different transverse size at freezeout, to each centrality bin, see Fig.~\ref{res01}.

We construct these centrality-dependent backgrounds by keeping $q$ fixed --- we choose $q=0.15$~fm$^{-1}$ --- and varying the value of $\hat T_0$
from  10.55 to 6.7 in steps of 0.55 for collisions with centralities of 5\%, 10\%, 20\%, 30\%, 40\%, 50\%, 60\% and 70\% --- see Table~\ref{table1} 
in Section~\ref{sec:Results}. If instead we had kept $\hat T_0$ fixed, changing the value of $q$ would
change the radius of the fireball described by the Gubser solution, but that is not the only consequence of changing $q$.  It is more straightforward to keep $q$ fixed and change $\hat T_0$, which keeps the shapes of isotherms in the Gubser solution unchanged but changes which isotherm in the solution is labeled with which temperature in MeV.

We have chosen the value $q=0.15$~fm$^{-1}$ and
the value $\hat T_0=8.9$ for mid-central (30\% centrality) 
collisions by
comparing the proton $p_T$-spectrum that 
we obtain in the next Section 
to experimental data from collisions with 20-30\% centrality, as shown in the next Section in Fig.~\ref{res02}. 

Choosing decreasing values of $\hat T_0$, meaning decreasing radii of the droplet of 
hydrodynamic fluid at freezeout, 
is a crude way to choose Gubser solutions with which to model collisions with increasing centralities.  It is not possible to do this realistically, since Gubser solutions are azimuthally symmetric while heavy ion collisions with nonzero impact parameter are not.  The choice that we have made (reducing $\hat T_0$ by 0.55 for each 10\% increase in the centralities of the collisions we model) balances the goals of describing both proton spectra and yields in more peripheral collisions.  
As we shall discuss further
in the next Section, 
with our choices for the values of $\hat T_0$ 
we overestimate the proton+antiproton yield
(and hence underestimate the baryon chemical potential $\mu_B$) in the most peripheral collisions.
In this regime, our choice of Gubser background yields proton spectra that are
too steep.
Addressing the first of these issues would require smaller values of $\hat T_0$ for the most peripheral collisions, whereas addressing the second of these issues would require larger values of $\hat T_0$. 
Notwithstanding these challenges that arise from using azimuthally symmetric Gubser solutions, 
the solutions that we employ have
values for the temperature at an initial time $\tau_i=0.6$~fm$/c$,
the effective initial transverse size, 
and the 
transverse size at
freezeout, that are not unreasonable: see
Fig.~\ref{res01}, and see Table~\ref{table1} 
in Section~\ref{sec:Results} for further details.

This strategy has clear limitations. Because the Gubser solution is azimuthally symmetric and has no explicit impact-parameter dependence, it cannot reproduce the full geometry of realistic noncentral collisions, especially in very central and very peripheral events. The centrality dependence of the electromagnetic field and the baryon-stopping profile is introduced through the impact parameter in those sectors of the calculation, while the hydrodynamic background accounts only effectively for the corresponding change in fireball size. For this reason, comparisons with data should not be interpreted as quantitative. 
Nevertheless, in the midcentral region where the sign change occurs, this set of centrality-dependent Gubser backgrounds provides a useful and internally consistent framework for studying the interplay between electromagnetic effects and baryon stopping.

\section{Freezeout procedure: single surface Cooper-Frye}
\label{sec:Freezeout}

We shall basically follow Ref.~\cite{Gursoy:2014aka} and  use the standard Cooper-Frye freezeout procedure \cite{PhysRevD.10.186} assuming sudden particlization and freezeout at the freezeout hypersurface
where the temperature 
in the hydrodynamic fluid has cooled to a specified freezeout temperature $T_f$.
We shall make no distinction between chemical and kinetic freezeout.

To set notation for the freezeout analysis, we write the  spectrum of hadron species \(i\) as
\begin{align}
S_i \;&\equiv\; p^0\frac{d^3N_i}{dp^3}
\;=\;
\frac{d^3N_i}{p_T\, dY\, dp_T\, d\phi_p} \\
\;&=\;
v_0\!\left[1+2v_1\cos(\phi_p-\pi)+2v_2\cos 2\phi_p+\cdots\right] ,
\label{eq:fourier_spectrum}
\end{align}
where \(Y\) is the momentum-space rapidity, \(p_T\) is the transverse momentum, and \(\phi_p\) is the azimuthal angle of the detected hadron momentum. In our convention, the spectators moving toward positive \(z\) (i.e. ``projectile'' spectators in Fig.~\ref{coll_geo}) are located at negative \(x\). Therefore, positive directed flow corresponds to deflection toward negative \(x\), which is why the first harmonic appears as \(\cos(\phi_p-\pi)\) rather than \(\cos(\phi_p)\). Equivalently, we define
\begin{align}
\label{eq:v0v1_defs}
v_0(p_T,Y)&=\frac{1}{2\pi}\int_{-\pi}^{\pi} d\phi_p\, S_i(p_T,Y,\phi_p),
\\
v_1(p_T,Y)
&=
\frac{\displaystyle \int_{-\pi}^{\pi} d\phi_p\, \cos(\phi_p-\pi)\, S_i(p_T,Y,\phi_p)}
{\displaystyle 2\pi\, v_0(p_T,Y)} . \nonumber
\end{align}

To convert our hydrodynamic background (before and at freezeout) into hadron spectra (at and after freezeout), we use the standard Cooper--Frye prescription on a single isothermal freezeout 
hypersurface specified via a choice
of freezeout temperature $T_f$
as $T(\tau,x_\perp)=T_f$.  
Accordingly, the hadron spectrum of species \(i\) is obtained from
\begin{equation}
S_i \;\equiv\; p^0\frac{d^3N_i}{dp^3}
\;=\;
-\frac{g_i}{(2\pi)^3}\int d\Sigma_\mu\, p^\mu\,
F_i(x,p) \, ,
\label{eq:CF_spectrum}
\end{equation}
where \(g_i\) is the degeneracy factor, \(d\Sigma_\mu\) is the directed area element of the freezeout hypersurface, and \(F_i(x,p)\) is the local phase-space distribution. 
We shall specify points on the freezeout surface by their proper time $\tau$, spacetime rapidity $\eta$, radial position $x_\perp$,
and azimuthal angle $\phi$.
For a hypersurface written as  $\Sigma^\mu=[\tau_f(x_\perp),x_\perp,\phi,\eta]$ where $\tau_f$ is defined 
via $T(\tau_f,x_\perp)=T_f$, the surface element is
\begin{align}
d\Sigma_\mu&=-\epsilon_{\mu\nu\alpha\beta}\frac{\partial\Sigma^\nu}{\partial x_\perp}\frac{\partial \Sigma^\alpha}{\partial \phi}\frac{\partial \Sigma^\beta}{\partial \eta} \sqrt{-g} d x_\perp d \phi d \eta \nonumber\\
&=
\left(
-1,\,
\partial \tau_f / \partial x_\perp,\,
0,\,
0
\right)
x_\perp \tau_f\,
d x_\perp d \phi d \eta.
\label{eq:CF_surface}
\end{align}

For the freezeout distributions $F_i$, we follow the same classical approximation used in \cite{Gursoy:2014aka} for and work directly with a classical Boltzmann distribution. 
This is sufficient for our purposes, since the main effect of interest comes from baryon stopping (that we shall turn on in Section~\ref{BScomp}) in concert with the radial flow in our hydrodynamic background 
and from electromagnetic fields (that we shall turn on in Section~\ref{EMcomp}), rather than from quantum-statistical corrections. 
However, we don't employ the simple Boltzmann distribution as Ref.~\cite{Gursoy:2014aka}, 
we upgrade it to the Maxwell--Boltzmann form in the presence of a chemical potential. 
The distribution then becomes
\begin{equation}
F_i(x,p)
=
\exp\!\left[\frac{p^\mu u_\mu+B_i\mu_B}{T_f}\right],
\label{eq:CF_distribution_muB_MB}
\end{equation}
where $B_i$ is the baryon number of hadron species $i$. 
This simple change is actually very important, as it will allow 
us to account for the net-proton asymmetry generated by baryon stopping
by introducing a spacetime-dependent baryon chemical potential on the freezeout surface. 
We present our explicit construction of \(\mu_B(x_\perp,\phi,\eta)\) in Section~\ref{BScomp}.

The physics encoded in the standard expression \eqref{eq:CF_distribution_muB_MB}
when $\mu_B$ varies across different regions of the freezeout surface can be thought of in two equivalent ways.
The fugacity factor $\exp(B_i\mu_B/T_f)$
weights the contribution of protons from different regions of the freezeout surface differently, depending on $\mu_B$. Equivalently, one can think
of the factor $\exp(p^\mu u_\mu/T_f)$
as encoding the central effect of radial flow that we wish to model: wherever on the freezeout surface $\mu_B$ is larger, reflecting the presence there of stopped baryons (protons in particular), 
this factor encodes the outward radial
boost that translates a rapidity-odd and reaction-plane-reflection-odd distribution of stopped protons into a rapidity-odd directed flow of protons.
 
Were we to neglect baryon stopping and assume that the baryon chemical potential $\mu_B=0$ vanished, the $\eta$- and $\phi_p$-integrated component \(v_0\) 
in the hadron spectrum 
\eqref{eq:fourier_spectrum}
could be written in analytic form because the  dependence is simple and the background remains fully symmetric. 
Once the \(\mu_B\)-dependent weight is included in $F_i$, however, this simplification is lost: the local distribution $F_i$ acquires additional spacetime structure, and the momentum integrals must be evaluated numerically. For this reason, in our implementation \(v_0\) is computed numerically in the same way that \(v_1\) must be computed. This is a small price to pay for incorporating the physically important baryon-stopping effect into an otherwise analytic freezeout framework.

\begin{figure}[t]
\centering
\includegraphics[width=1.\columnwidth]{"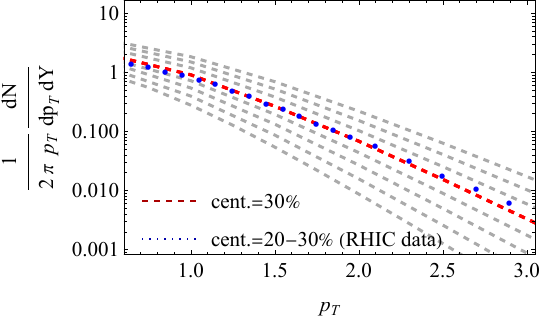"}
\caption{Comparison of the proton $p_T$ spectrum obtained from the Gubser solution with parameters that we employ (chosen as described in the text) to model mid-central heavy ion collisions with 20-30\% centrality, shown by the red dashed curve, with the data from Ref.~\cite{PHENIX:2003iij}, shown by the blue dots. 
The gray dashed curves show the 
proton spectra obtained from the Gubser solutions that we employ to model heavy ion collisions with other centralities, with the gray curves from top to bottom running from central to peripheral as in Fig.~\ref{res01}.
Our model for 30\% centrality collisions provides a good fit to the data. 
We have chosen parameters so as to reproduce the normalization of the
proton spectrum for mid-central collisions
as these collisions will be our focus.
}
\label{res02}
\end{figure}

Because the pure Gubser background (with $\mu_B=0$ at freezeout and with no
electromagnetic field perturbations)
is boost invariant and azimuthally symmetric, the hadron spectra that result in this case are independent of both \(Y\) and \(\phi_p\). 
As a result, all anisotropic coefficients vanish in pure Gubser flow, \(v_{n\geq 1}=0\), and only the isotropic component \(v_0\) remains nonzero.  We must therefore use results for $v_0$ to fix the freezeout temperature $T_f$ and the 
parameters $q$ and $\hat T_0$ that specify the Gubser solution.  We 
shall take $T_f=160$~MeV; we
have specified our choices for $q$ and $\hat T_0$ in the previous Section.
We motivate these choices here, via the comparison in Fig.~\ref{res02}.

%%%%%%%%%%%%%%%%%%%%%

Our choice of $T_f$ is comparable to
the standard chemical freezeout temperature. This makes sense given that
we will be specifying the density of 
stopped baryons via the fugacity factor $\exp(B_i\mu_B/T_f)$ in the Cooper--Frye distribution at the freezeout temperature $T_f$.  It is perhaps coincidental that when we use the Gubser solution as our hydrodynamic background and make this reasonable choice of $T_f$, if we make the (unreasonable) assumption that kinetic freezeout occurs at the same temperature $T_f$ we obtain a reasonable description of the slope of the proton $p_T$ spectrum, see Fig.~\ref{res02}.
In a future calculation with a realistic hydrodynamic background, one will need to introduce a separate, lower, kinetic freezeout temperature in order to obtain a reasonable description of hadron spectra.

%%%%%%%%%%%%%%%%%%%%%%%%%%%%%%%%%%%%%%
\section{Implementing Baryon Stopping}\label{BScomp} 

Baryon stopping refers to the loss of longitudinal momentum by a fraction of the incoming baryons in heavy-ion collisions. Instead of continuing close to beam rapidity, these baryons deposit baryon number and energy into the interaction region, thereby generating a nonzero net-baryon density around mid-rapidity. In this sense, baryon stopping is a mechanism that transports some of the conserved baryon number over a large rapidity interval and results in the production of a baryon-doped droplet of quark-gluon plasma. For the present work, the main consequence of the fact that the fireball no longer has zero net baryon number is that proton-rich regions can appear on the freezeout surface, and these regions can affect the charge-dependent directed flow measured in the final state.
We will in particular be interested in the component of the distribution of stopped baryons in a collision with nonzero impact parameter that is both odd in rapidity and odd under reflection in the impact parameter axis $x$, see Fig.~\ref{coll_geo}, as radial flow turns this component of the stopped proton distribution into a contribution the charge-dependent directed flow.  
To our knowledge, the charge-dependent directed flow is the only observable that has been measured in experimental data that has a significant
sensitivity to this component of the stopped baryon distribution. This makes analyses of this observable, such as ours, interesting as diagnostics of how well we understand baryon stopping as well as for revealing electromagnetic effects.

We encode the effect of the accumulated stopped baryon number through a local baryon chemical potential \(\mu_B(x_\perp,\phi,\eta)\).  
Because \(\mu_B\) distinguishes baryons from antibaryons in the freezeout distribution, together with the fact that the baryon-rich matter resulting from 
baryon stopping is subsequently carried outward by the radial expansion of the hydrodynamic droplet it serves to introduce a 
splitting between proton and antiproton directed flow. We shall see that, as in Refs.~\cite{Nayak:2019vtn,Guo:2012qi,Bozek:2022svy}, this plays a vital role in the rapidity-slope of the charge-dependent directed flow, $d \Delta v_1(p)/dy$.

\subsection{Local Baryon Chemical Potential}
For RHIC collisions at \(\sqrt{s_{NN}}=200~\mathrm{GeV}\), 
analyses of experimental data on 
hadron multiplicities indicate that
the average baryon chemical potential is
$\mu_B\sim 22$~MeV (for example, see 
Ref.~\cite{Andronic:2008gu}) meaning $\mu_B\ll T_f$.
We shall assume that this inequality is
 well-satisfied everywhere on the freezeout surface.
 This means that in our construction of the local chemical potential $\mu_B(x_\perp,\phi,\eta)$ on the freezeout hypersurface, 
it is consistent to work in the small-\(\mu_B\) regime and to use the linearized relation between the local net-baryon density and the local baryon chemical potential:
\begin{equation}
\mu_B(x_\perp,\phi,\eta)\approx \frac{n_B(x_\perp,\phi,\eta)}{\chi_B(T_f)} \, ,
\label{eq:muB_from_nB_rewrite}
\end{equation}
where \(\chi_B(T)\) is the baryon susceptibility. To fix the baryon susceptibility, we use the results from the HotQCD calculations of lattice QCD thermodynamics in  Ref.~\cite{Bazavov:2017dus}, which yield a dimensionless susceptibility $\chi_B/T^2 \simeq 0.12$ at $T\simeq 160~{\rm MeV}$. 

\subsection{Constructing the Stopped Baryon Density in the Glauber Model}\label{gb-section}

The remaining question is how to model \(n_B(x_\perp,\phi,\eta)\). In the present framework, \(n_B\) should be understood as the net baryon density deposited by baryon stopping. We shall model this quantity  using the participant density from a standard Glauber-model description of the nuclear geometry, described below, since the stopped baryons originate from participant nucleons and their spatial distribution should therefore follow the participant profile. Accordingly, we take
\begin{equation}
n_B(x_\perp,\phi,\eta)\propto n_{\mathrm{part}}(x_\perp,\phi,\eta) \, .
\label{eq:nB_npart_rewrite}
\end{equation}
Because \(n_{\mathrm{part}}\) is a density per unit area in the transverse plane  in the Glauber model whereas \(n_B\) is a volume density, 
we convert the former into the latter via
\begin{equation}
n_B(x_\perp,\phi,\eta)=\frac{C_B}{\tau_f}\,n_{\mathrm{part}}(x_\perp,\phi,\eta) \, ,
\label{eq:nB_conversion_rewrite}
\end{equation}
where \(C_B\) is a dimensionless geometric conversion factor 
%(to convert the areal density to the  volume density) 
and $\tau_f(x_\perp)$ 
is the local proper time on the freezeout surface. In this way, the Glauber participant distribution provides a simple proxy for the spacetime pattern of stopped baryons, which is then mapped into the local baryon chemical potential through Eq.~\eqref{eq:muB_from_nB_rewrite}. In this subsection, 
we explicitly construct $n_{part}(x_\perp,\phi,\eta)$ in the Glauber model.

To fix $C_B$, we use a simple phenomenological criterion based on the measured proton--antiproton asymmetry in Au+Au collisions at $\sqrt{s_{NN}}=200$~GeV at RHIC. This asymmetry is quantified by the ratio $\bar p/p$, which is approximately $0.8$ and shows only a mild centrality dependence~\cite{PHENIX:2003iij}. For simplicity, we fix $C_B$ using midcentral events and then keep the same value for all centrality bins. 
The choice $C_B=12.85$ gives $\bar p/p \approx 0.8$ in midcentral collisions with 20-30\% centrality, which is the criterion that we use to fix the overall normalization of the stopped-baryon density. 
As we are anyway not able to do a realistic phenomenological analysis using Gubser hydrodynamic solutions, for simplicity we have chosen to keep  
the same value of $C_B$ for all centralities.
We tabulate the $\bar p/p$ as a function of centrality in our model in Sec.~\ref{sec:Results}. 
With the choices we have made here and in the previous Section, in our model the $\bar p/p$ ratio 
rises with increasing centrality in a way that is not seen in experimental data.
There are two reasons for this. The first is that with the choices of parameters specifying the Gubser solutions that we employ to model collisions with increasing centrality, as we discussed in the previous Section our model 
has too large a proton+antiproton yield in the most peripheral collisions.  The second is that choosing a single constant value of $C_B$ means that since the number of participants per unit transverse area is greatest in head-on collisions, so too is the stopped-baryon density.
We note, however, that the ratio $\bar p/p$ 
is not itself important for our principal calculation and does not play an important role in any of our conclusions:
it correponds to the average $\mu_B$ across the entire freezeout surface, whereas in this paper what we are interested in is
the spatial distribution of stopped protons. 
Specifically, 
we are interested in the rapidity-odd and reflection-odd component of the stopped proton distribution, as we have discussed around Fig.~\ref{coll_geo}. The multiplicity ratio $\bar p/p$ tells us nothing about this component, which vanishes when averaged across the entire freezeout surface. We shall confirm, though, that the charge-dependent  proton directed flow $v_1(p)-v_1(\bar p)$ observable, the analysis of which is our goal, is indeed sensitive to this feature
of the stopped proton distribution.

To construct the spatial distribution of stopped baryons, we use a standard Glauber-model description of the nuclear geometry. In this framework, both $n_{\rm part}$ and particle production and are determined by the transverse overlap of the two incoming nuclei at fixed impact parameter \(b\). The advantage of this approach is that it provides a simple and physically transparent way to encode the asymmetric geometry of a non-central collision while keeping the calculation analytically manageable.

In the Glauber framework, the basic input is the nuclear density profile. For a spherical nucleus we take the Woods--Saxon form
\begin{equation}
\rho_{3D}(r)=\frac{\rho_0}{1+\exp\!\left(\dfrac{r-R_A}{a_0}\right)},
\qquad
r=\sqrt{s^2+z^2},
\label{eq:WS}
\end{equation}
where $A$ is the atomic mass number and where $s=(x,y)$ or, with polar coordinates for the transverse plane, $s=(x_\perp,\phi)$. The parameter $\rho_0$ is an overall constant that is fixed by normalization of the distribution.  \(R_A\) is the nuclear radius parameter, and \(a_0\) is the surface thickness parameter. For Au, these parameters are as follows: $A=197$, $R_A=6.38$ fm and $a_0=0.535$ fm. 
Integrating over the longitudinal direction gives the standard thickness function
\begin{equation}
T_A(s)\equiv \int_{-\infty}^{+\infty} dz\,
\rho_{3D}\!\left(\sqrt{|s|^2+z^2}\right),
\quad
\int d^2s\, T_A(s)=1,
\label{eq:TA}
\end{equation}
which has the interpretation of a transverse probability density for finding a nucleon.

\begin{figure}[t]
\centering
\includegraphics[width=0.9\columnwidth]{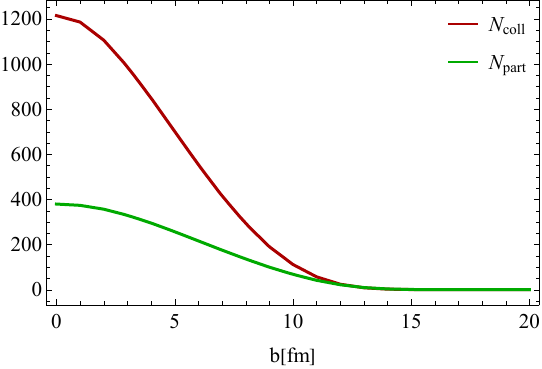}
\caption{Impact-parameter dependence of the total number of participants $N_{\rm part}(b)$ and binary nucleon--nucleon collisions $N_{\rm coll}(b)$ obtained from the optical Glauber model for Au+Au collisions at $\sqrt{s_{NN}}=200~{\rm GeV}$. These quantities define the geometric input used to construct the stopped-baryon profile and to assign centrality-dependent collision geometry in our calculations of baryon-stopping and electromagnetic fields.}
\label{npart}
\end{figure}

With the conventions of Fig.~\ref{coll_geo}, the 
forward-going nucleus (P in Fig.~\ref{coll_geo}) is centered at \(x=-b/2\), while the backward-moving nucleus (T in Fig.~\ref{coll_geo}) is centered at \(x=+b/2\). Accordingly, we define
\begin{equation}
T^\pm(s;b)=T_A\!\left(s-s_\pm\right),
\qquad
s_\pm=\left(\mp \frac{b}{2},\,0\right).
\label{eq:Tpm}
\end{equation}
From these shifted thickness functions we construct the participant thickness functions,
\begin{align}
T^\pm_{\rm part}(s;b)&=T^\pm(s;b)\Bigl(1-\bigl[1-\sigma_{NN}T^\mp(s;b)\bigr]^A\Bigr),
\label{eq:Tpartpm}
\end{align}
where \(\sigma_{NN}\) is the inelastic nucleon--nucleon cross section. For Au+Au collisions at $\sqrt{s_{NN}}=200~\mathrm{GeV}$, we use the standard value
$\sigma_{NN}=42~\mathrm{mb}$~\cite{Miller:2007ri}.
These quantify the transverse densities of nucleons from the forward- and backward-going nuclei that suffer at least one inelastic collision.
The corresponding total number of participants and the number of binary collisions (that can yield particle production) are
\begin{align}
N_{\rm part}(b)
&=
A\left[
\int d^2s\, T^+_{\rm part}(s;b)
+
\int d^2s\, T^-_{\rm part}(s;b)
\right],\\
N_{\rm coll}(b)&=A^2\sigma_{NN}T_{AA}(b),
\label{eq:Npart}
\end{align}
where  
\begin{equation}
T_{AA}(b)=\int d^2s\, T^+(s;b)\,T^-(s;b).
\label{eq:Ncoll}
\end{equation}
The resulting \(N_{\rm part}(b)\) and \(N_{\rm coll}(b)\) curves, shown in Fig.~\ref{npart}, are consistent with the results of Refs.~\cite{Kharzeev:2000ph,Miller:2007ri}.

\begin{figure}[t]
\centering
\includegraphics[width=0.9\columnwidth]{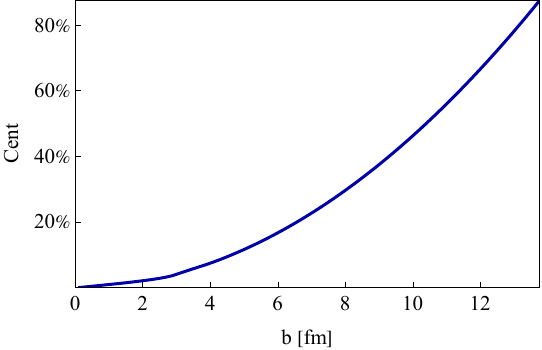}
\caption{Relation between the impact parameter $b$ and the centrality percentile $C(b)$ obtained from the inelastic Au+Au cross section in the optical Glauber model. This mapping is used throughout the calculation to convert the centrality classes used in the data comparison into the corresponding impact parameters entering the baryon-stopping and electromagnetic-field profiles.}
\label{cofb}
\end{figure}

To relate the geometry to centrality, we use the inelastic nucleus--nucleus cross section
\begin{equation}
\sigma_{AA}
=
\int_0^\infty 2\pi b' \, db'\,
\left[
1-\bigl(1-\sigma_{NN}T_{AA}(b')\bigr)^{A^2}
\right],
\label{eq:sigmaAA}
\end{equation}
and define the cumulative fraction up to impact parameter \(b\) as
\begin{equation}
C(b)=\frac{1}{\sigma_{AA}}
\int_0^b 2\pi b' \, db'\,
\left[
1-\bigl(1-\sigma_{NN}T_{AA}(b')\bigr)^{A^2}
\right].
\label{eq:Cofb}
\end{equation}
This provides the map between centrality percentile and impact parameter used throughout our analysis plotted in Fig.~\ref{cofb}, which is 
consistent with the results of Refs.~\cite{Kharzeev:2000ph,Broniowski:2001ei}. 

We next introduce the rapidity dependence of the stopped-baryon profile, which is critical for our purposes. We employ the empirical distribution \cite{KHARZEEV1996238,KHARZEEV2008227}
\begin{equation}
f^\pm(\eta)=\frac{a}{2\sinh(aY_0)}\,e^{\pm a\eta},
\qquad
-Y_0\le \eta \le Y_0 .
\label{eq:feta_BS}
\end{equation}
where $+$ ($-$) is for the forward (backward) moving participants and $Y_0$ is the rapidity of the incident nuclei in the center-of-mass frame. In Regge theory, the parameter $a$ is the string junction exchange intercept~\cite{KHARZEEV1996238}; its value is consistent with experimental 
data for baryon stopping~\cite{KHARZEEV1996238,ALICE:2013yba}. We use $a=0.82$ which is the most recent updated value~\cite{Frenklakh:2024mgu}. 
Note that the distribution 
in Eq.~\eqref{eq:feta_BS} was
originally written as a distribution in momentum-space rapidity $Y$.~\cite{KHARZEEV1996238,KHARZEEV2008227}. Because the Gubser solution that we are employing is boost invariant, with $Y=\eta$, it is natural to use the same profile as a function of spacetime rapidity $\eta$.

With this longitudinal profile, one can define a baryon-density ansatz directly from the thickness functions,
\begin{equation}
n_B(s,\eta;b)
=
A\Bigl[
T^+(s;b)\,f^+(\eta)+T^-(s;b)\,f^-(\eta)
\Bigr].
\label{eq:nB_profile}
\end{equation}
However, for modeling baryon stopping it is more appropriate to use the participant thickness functions, since the stopped baryons are associated with nucleons that actually interact. We therefore define
\begin{equation}
n_{\rm part}(s,\eta;b)
=
A\Bigl[
T^+_{\rm part}(s;b)\,f^+(\eta)+T^-_{\rm part}(s;b)\,f^-(\eta)
\Bigr].
\label{eq:npart_profile}
\end{equation}
This is the profile that we shall use  in our implementation of baryon stopping. 
It combines the transverse geometry of the participant matter with a longitudinal distribution inherited from baryon stopping, which automatically satisfies the $A{+}A$ symmetry
$n_{\rm part}(x,y,\eta;b)=n_{\rm part}(-x,y,-\eta;b)$.  This is precisely the rapidity-odd and reflection odd component of the stopped baryon distribution that we foreshadowed in Fig.~\ref{coll_geo}
that is needed in order for the radial flow of the stopped protons 
%proton stopping and radial flow 
to induce a rapidity-odd contribution to charge-dependent directed flow.

In summary, the Glauber model provides the geometric backbone for our construction of the stopped-baryon density. The transverse overlap determines where participant matter is deposited, while the rapidity weights \(f^\pm(\eta)\) distinguish the contributions from the two beam directions. The resulting profile \(n_{\rm part}(x,y,\eta;b)\) then serves as the proxy for the distribution of stopped baryons that will subsequently be mapped onto the local baryon chemical potential via Eqs.~\eqref{eq:nB_conversion_rewrite} 
and \eqref{eq:muB_from_nB_rewrite}.
This local baryon chemical
potential encodes the distribution of
stopped baryons on the freezeout surface and, via the distribution \eqref{eq:CF_distribution_muB_MB}
that governs Cooper-Frye freezeout,
the interplay between this rapidity-odd and reflection-odd stopped proton distribution and the radial flow of the Gubser solution yields a rapidity-odd contribution to the charge-dependent directed flow of protons.

%%%%%%%%%%%%%%%%%%%%%%%%%%%%%%%%%%%
%%%%%%%%%%%%%%%%%%%%%%%%%%%%%%%%%%%

\section{CONSTRUCTING THE ELECTROMAGNETIC FIELD}
\label{EMcomp}
%%%%%%%%%%%%%%%%%%%%%%%%%%%%%%%%%%%
%%%%%%%%%%%%%%%%%%%%%%%%%%%%%%%%%%%
In this Section, we describe the construction of the electromagnetic field used in our calculation. We follow the semi-analytic framework of Ref.~\cite{Gursoy:2014aka}, where the electromagnetic field generated by the charged nucleons is obtained by solving Maxwell equations in a conducting medium. This framework was later implemented in realistic hydrodynamic simulations in Ref.~\cite{Gursoy:2020jso}. We also note that similar ideas were developed in earlier studies of the time evolution of electromagnetic fields in heavy-ion collisions~\cite{Skokov:2009qp,Tuchin:2010vs,Voronyuk:2011jd,Deng:2012pc,Tuchin:2013ie,McLerran:2013hla}. Since our goal is only to specify the electromagnetic input used in the force-balance calculation that we shall present below
that quantifies the charge-dependent drift caused by electromagnetic fields, we summarize the basic ingredients of the framework and refer the reader to Ref.~\cite{Gursoy:2014aka} for the full derivation.

We start by identifying the field components relevant for the directed-flow calculation. With the collision geometry shown in Fig.~\ref{coll_geo}, where the beam is along the $z$ direction and the impact parameter is along the $x$ direction, the dominant component of the magnetic field in the center-of-mass frame is $B_y$. The corresponding transverse electric field $E_x$ is the component that produces a sideward force along the reaction plane, which can drive a charge-dependent directed flow. Other components of the magnetic and electric fields are not included in the present calculation: they either vanish by symmetry in the smooth geometry, are fluctuation-driven, or do not contribute to the leading rapidity-odd sideward response considered here~\cite{Gursoy:2014aka,Gursoy:2020jso,Dubla:2020bdz}. Thus, the electromagnetic fields that enter the charge-dependent drift calculation are %taken to be
\begin{equation}
    \mathbf{B} \simeq B_y\,\hat{\mathbf{y}},
    \qquad
    \mathbf{E} \simeq E_x\,\hat{\mathbf{x}} \ ,
\end{equation}
which we need to determine as a function of position and time.

As a simplifying assumption, we neglect the participant contribution to the electromagnetic field. This is consistent with the simplified nature of the present calculation and with the observation that the dominant contribution to the rapidity-odd directed-flow signal comes from the spectator-induced fields. The participant contribution is substantially smaller and changes the final result by at most about $10\%$, and typically by much less~\cite{Gursoy:2014aka}.

Following Ref.~\cite{Gursoy:2014aka}, we first compute the field generated by a single point charge moving in the beam direction with some nonzero impact parameter. This requires solving 
the Maxwell equations in a conducting 
medium for 
the magnetic and electric fields sourced by the point charge.
Once $B_y$ and $E_x$ are obtained for a single charge, the total field is constructed by summing the single-particle solution over the spectator protons from the two nuclei. In the continuous-density approximation, this sum is replaced by an integral over the transverse spectator density. In Sec.~\ref{sec:ComputingDirectedFlow}, we shall use the resulting electromagnetic field as an input in the force-balance equation via which we compute the charge-dependent directed flow.

\begin{figure}[t]
\centering
\includegraphics[width=1.\columnwidth]{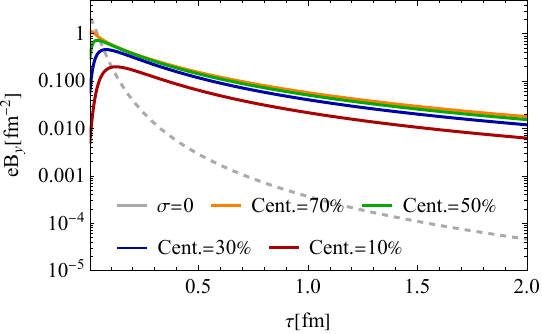}
\caption{Time dependence of the spectator-induced magnetic field $eB_y$ at the center of the collision, $\eta=x_\perp=0$, for Au+Au collisions at $\sqrt{s_{NN}}=200~{\rm GeV}$. The solid curves show results for different centrality classes with a constant nonzero electrical conductivity $\sigma=0.032$~fm$^{-1}$ as described in the text, while the gray dashed curve shows the corresponding result in the limit $\sigma=0$. The magnetic field increases toward more peripheral collisions due to the larger spectator contribution; the nonzero conductivity slows its decay in time.}\label{EM}
\end{figure}

In Fig.~\ref{EM}, we show the time-dependence of the spectator-induced magnetic field $eB_y$ at the center of the collision, $\eta=x_\perp=0$, for Au+Au collisions at $\sqrt{s_{NN}}=200~{\rm GeV}$. The solid curves correspond to different centrality classes. As expected, the magnitude of $eB_y$ increases from central to peripheral collisions, reflecting the larger spectator contribution at larger impact parameter. The gray dashed curve shows the corresponding result in the limit of vanishing electrical conductivity, included for comparison. The difference between the dashed and solid curves illustrates the role of the conducting medium in slowing the decay of the magnetic field.

The electrical conductivity $\sigma$ controls the time evolution of the electromagnetic field in the conducting medium and, consequently, the relative size of the Faraday and Lorentz contributions to the charge-dependent directed flow. 
We shall follow
Refs.~\cite{Gursoy:2014aka,Gursoy:2018yai}
and 
treat $\sigma$ as 
if it were a constant as this substantially simplifies the calculation, leaving taking account of its temperature-dependence and consequent variation with space and time to a future magnetohydrodynamic analysis.
As in Refs.~\cite{Gursoy:2014aka,Gursoy:2018yai},
we choose a value of $\sigma$ based upon lattice QCD calculations of 
$C_{\rm em}^{-1}\sigma/T$, where $C_{\rm em}\equiv\sum_i q_i e^2=0.061$, with the sum taken over quark flavors $i=u,d,s$. 
Based upon the results from unquenched lattice QCD calculations of 
$C_{\rm em}^{-1}\sigma/T$ at $T=350$~MeV (comparable to the initial temperature of the QGP that forms in a RHIC
collision and traps some of the initial magnetic field)
done with either 2 or 2+1 flavors of dynamical quarks~\cite{Brandt:2012jc,Amato:2013naa,Aarts:2014nba,Brandt:2015aqk,Astrakhantsev:2019zkr}, reviewed in Ref.~\cite{Aarts:2020dda},
we shall fix $\sigma=0.032$~fm$^{-1}$.
This is somewhat larger than the value used in Refs.~\cite{Gursoy:2014aka,Gursoy:2018yai},
which was based upon results from unquenched lattice 
calculations~\cite{Ding:2010ga,Francis:2011bt,Ding:2016hua}
at the lower temperature $1.5~T_c$.

%%%%%%%%%%%%%%%%%%%%%%%%%%%%%%%%%%%
%%%%%%%%%%%%%%%%%%%%%%%%%%%%%%%%%%%
\section{COMPUTING DIRECTED FLOW INDUCED BY BARYON STOPPING AND ELECTROMAGNETIC FIELDS}
\label{sec:ComputingDirectedFlow}
Our calculation follows the same perturbative strategy as in Refs.~\cite{Gursoy:2014aka,Gursoy:2018yai}. The basic idea is to separate the bulk hydrodynamic expansion from the much smaller charge-dependent drift generated by the electromagnetic field. After  determining the spacetime-dependent electric and magnetic fields produced in a collision with nonzero impact parameter (see Sec.~\ref{EMcomp}), in this Section we evaluate the small velocity induced on charged carriers by these fields in the expanding medium. 
This induced drift is added as a small correction on top of the background flow, after which the resulting hadron spectra and flow harmonics are obtained from the Cooper--Frye 
freezeout prescription.  As we have described in Secs.~\ref{sec:Freezeout} and \ref{BScomp}, the contribution from the stopped protons, boosted by radial flow, enters via the freezeout calculation.

Concretely, after constructing the electromagnetic field for a collision at fixed impact parameter, we determine the induced drift velocity for charged constiuents in the medium
in the same perturbative manner as in Refs.~\cite{Gursoy:2014aka,Gursoy:2018yai}. 
Since this drift is assumed to remain much smaller than the background hydrodynamic flow, it is sufficient to work in the local fluid rest frame and solve a stationary force-balance equation in that frame. For a charged fluid element of charge \(q\) and effective mass \(m\), the drift velocity \(\vec v^{\,\prime}\) is obtained from
\begin{equation}
m\frac{d\vec v^{\,\prime}}{dt}
=
q\,\vec v^{\,\prime}\times \vec B^{\,\prime}
+
q\,\vec E^{\,\prime}
-
\mu m\,\vec v^{\,\prime}
=0 \, ,
\label{eq:force_balance_section}
\end{equation}
where \(\vec E^{\,\prime}\) and \(\vec B^{\,\prime}\) are the electromagnetic fields in the local fluid rest frame. The last term represents the drag exerted by the medium and opposes the Lorentz force. The use of the nonrelativistic form of Eq.~\eqref{eq:force_balance_section} is justified precisely by the perturbative assumption that the induced drift is small compared to the background flow. 

In Refs.~\cite{Gursoy:2014aka,Gursoy:2018yai}, the drag parameter $\mu$ was estimated 
starting from the result from holographic calculations of the drag force on heavy quarks in the strongly coupled plasma of ${\cal N}=4$ supersymmetric Yang-Mills (SYM) theory, for which~\cite{Herzog:2006gh,Gubser:2006bz}~
\begin{equation}
\mu m=\frac{\pi\sqrt{\lambda}}{2}\,T^2 \, 
\label{eq:drag_section}
\end{equation}
in the limit of a large number of colors and large 't Hooft coupling  \(\lambda=g^2N_c\).  
The authors of Refs.~\cite{Gursoy:2014aka,Gursoy:2018yai} employed Eq.~\eqref{eq:drag_section} with $\sqrt{\lambda}=6\pi$ to evaluate $\mu m$ in the force-balance equation~\eqref{eq:force_balance_section}.  By now it is clear that this would be an overestimate for the drag force on a heavy quark in QGP, as strongly coupled QGP in QCD has fewer degrees of freedom (meaning a smaller energy and entropy density at the same temperature and coupling) than
the strongly coupled ${\cal N}=4$ SYM plasma. Following Ref.~\cite{Beraudo:2025nvq}, we rewrite
Eq.~\eqref{eq:drag_section} as
\begin{equation}
\mu m=\frac{\kappa_{HQ}}{2}\,T^2 \ , 
\label{eq:drag-with-kappaHQ}
\end{equation}
understanding that $\kappa_{HQ}$ is a parameter whose value for heavy quarks being dragged through the QGP of QCD should be determined by comparison between model calculations and experimental measurements of heavy hadron spectra and flow in heavy ion collisions. The authors of 
Ref.~\cite{Beraudo:2025nvq} find reasonable descriptions of current data, within uncertainties, for $3.4 \leq \kappa_{HQ} \leq 6.3$ and quote most of their results for $\kappa_{HQ}=4.4$, which corresponds to a value of $\mu m$ roughly $1/3$ of that used 
Refs.~\cite{Gursoy:2014aka,Gursoy:2018yai}.  Noting that we are employing 
Eq.~\eqref{eq:drag-with-kappaHQ} in a way for which it was not designed, as we are discussing the drag on a charged fluid element with effective mass $m$ and not a heavy quark, meaning that the uncertainties here are even greater than those in 
Ref.~\cite{Beraudo:2025nvq}, we shall take $\mu m$ to be $1/3$ of that used 
Refs.~\cite{Gursoy:2014aka,Gursoy:2018yai}, which corresponds to choosing $\kappa_{HQ}=4.55$.

Ultimately, the parameter $\mu m$
in the force-balance equation \eqref{eq:force_balance_section}
should be determined by comparing calculations of charge-dependent directed flow (like ours, but done with a realistic hydrodynamic background)
to experimental data.
At present, our lack of knowledge of the values of $\mu m$ and the electrical conductivity $\sigma$ (discussed in the previous Section) are
two important sources of uncertainty in our modeling, as these are
the two 
transport properties of QGP 
on which the charge-dependent directed flow that we calculate depends. 

As in Refs.~\cite{Gursoy:2014aka,Gursoy:2018yai},
we solve the force-balance equation \eqref{eq:force_balance_section} separately for each quark flavor, since the electromagnetic force depends on the charge. 
We compute the drift velocities for \(u\), \(d\), \(\bar u\), and \(\bar d\) quarks individually in the local fluid rest frame, with each picking up a drift velocity in this frame proportional to its electric charge: \(q_u=+2e/3\), \(q_{d}=-e/3\), \(q_{\bar u}=-2e/3\) and  \(q_{\bar d}=+e/3\). 
Following Refs.~\cite{Gursoy:2014aka,Gursoy:2018yai}, we neglect strange quarks and assume that
in the strongly coupled hydrodynamic fluid before freezeout the 
average drift velocity of positive charges is proportional to $+e/2$ and the average drift velocity of negative charges is proportional to $-e/2$, and assign these drift velocities to
positively and negatively charged hadrons as they form at the freezeout surface.
In this way, our freezeout prescription
maps an electric current in the hydrodynamic fluid immediately before freezeout (induced by the electromagnetic forces that we have analyzed) onto the same charge-current in the gas of hadrons immediately after freezeout.
After obtaining the drift velocities of 
charged hadrons at the freezeout surface
in the local fluid rest frame, we
boost these drift velocities back to the lab frame using the local hydrodynamic velocity from the Gubser solution,
yielding the full charge-dependent four-velocity \(u_\mu\). 
This yields a total flow field that contains the dominant hydrodynamic expansion plus a small charge-dependent correction. Finally, this 
corrected flow $u_\mu$
is inserted into the 
distribution $F_i$ (see Eq.~\eqref{eq:CF_distribution_muB_MB})
that appears in the Cooper-Frye
freezeout integral \eqref{eq:CF_spectrum},
allowing us
to compute the directed flow for each hadron species using Eq.~\eqref{eq:v0v1_defs}. 
Since the unperturbed Gubser background is azimuthally symmetric, it gives \(v_{n\geq 1}=0\) by itself; thus the nonzero directed flow \(v_1\) obtained via this procedure directly reflects the perturbing effects that break this symmetry.

Charge-dependent directed flow originates from two physical effects that the procedure we have described treats together, consistently.
One effect enters via solving the
force-balance equation \eqref{eq:force_balance_section} 
in the presence of electromagnetic fields: hadrons at the freezeout surface with opposite charges pick up small drift velocities in opposite 
directions. With the geometry that we described in Fig.~\ref{coll_geo}, this yields a charge-dependent rapidity-odd directed flow.
The second effect arises because,
as we described in Sec.~\ref{BScomp},
the distribution of stopped baryons is
rapidity-odd and reflection-odd, as reflected in
our calculation by the contribution to the
local baryon chemical potential \(\mu_B(x,\phi,\eta)\)
on the freezeout surface that is odd under both $\eta\rightarrow -\eta$ and $x\rightarrow -x$. 
Via Eq.~\eqref{eq:CF_distribution_muB_MB},
the freezeout distribution 
is modified by the local factor \(e^{B_i\mu_B/T_f}\), so baryons and antibaryons are weighted differently, with the difference between protons and antiprotons yielding a charge-dependent directed flow.
Physically, this incorporates the fact that a nonzero density of stopped protons accumulates in a rapidity-odd fashion, with opposite signs in opposite directions along the impact parameter axis $x$ (see Fig.~\ref{coll_geo}), with this excess of positively charged protons
then being carried outward by the radial expansion of the medium described in our calculation by the Gubser solution, namely the unperturbed $u_\mu$.
The electromagnetic field and baryon stopping both enter our calculation of the charge-dependent directed flow 
via the consistent calculation of the
distribution $F_i$ in Eq.~\eqref{eq:CF_distribution_muB_MB},doing so
through distinct but complementary 
mechanisms: the former generates a charge-dependent drift velocity, a small charge-dependent 
perturbation to $u_\mu$, while the latter  changes the local baryon versus antibaryon emission weight on the freezeout surface via $\mu_B$ and relies upon the large unperturbed $u_\mu$. 

We model the centrality dependence by repeating this construction at different impact parameters. As we have discussed in Sec.~\ref{sec:HydroBackground},
we use different Gubser solutions (with different radii) to model collisions with different impact parameter $b$.  
And, both the electromagnetic field and the baryon chemical potential are recalculated for each \(b\). The magnetic and electric fields depend explicitly on the collision geometry, with their strength increasing with increasing $b$ as there are more spectators in more peripheral collisions. The local baryon chemical potential \(\mu_B(x,\phi,\eta)\) depends on the Glauber participant distribution 
which depends on $b$, with
the mapping between impact parameter and centrality provided by the cumulative Glauber function \(C(b)\) from Eq.~\eqref{eq:Cofb} plotted in Fig.~\ref{cofb}.  
Thus, the centrality dependence of our results comes directly from the impact-parameter dependence of the electromagnetic field and of the stopped-baryon profile.

We note that we expect the model to become less reliable both in very central
collisions and in very peripheral collisions. 
In very central collisions, the
electromagnetic field that we calculate is small by symmetry, while in reality the directed flow can receive important contributions from event-by-event fluctuations in the shape of the collision that we do not model.
In very peripheral collisions, the model is also less reliable because the 
medium produced in such collisions
is smaller and shorter lived, and for both reasons the application of hydrodynamics to such collisions becomes more challenging and fluctuations that we do not model also become important.
For these reasons, the most meaningful comparisons between the qualitative trends in our results and  in experimental data are those
in the intermediate centrality range, where both a sizable medium and sizable electromagnetic fields are present, while the basic assumptions of the framework remain qualitatively reasonable.

%%%%%%%%%%%%%%%%%%%%%
\begin{table*}[t]
\centering
\begin{tabular}{|c|c|c|c|c|c|c|c|c|c|c|c|}
\hline
Centrality & $\hat T_0$ & $q$ & $C_B$ & $\sigma$ & $\mu m$ & $\tau_f(x_\perp=0)$ & $R_\perp^{f}$ &
$\langle T_i^{(c)}\rangle$& $\langle T_i\rangle$ & $\bar{p}/p$ & $\langle \mu_B^f \rangle$  \\
\hline
5\% & 10.55 & $0.15~\mathrm{fm}^{-1}$ & 12.85 & 0.032 $~\mathrm{fm}^{-1}$ & $4.55\,T^2/2$ & 7.1 fm$/c$ & 14.8 fm & 585 MeV& 485 MeV & 0.70 & 29 MeV \\
\hline
10\% & 10.0 & $0.15~\mathrm{fm}^{-1}$ & 12.85 & 0.032 $~\mathrm{fm}^{-1}$ & $4.55 \, T^2/2$ & 6.76 fm$/c$ & 14.1 fm & 553 MeV& 458 MeV & 0.74 & 24 MeV\\
\hline
20\% & 9.45 & $0.15~\mathrm{fm}^{-1}$ & 12.85 & 0.032 $~\mathrm{fm}^{-1}$ & $4.55 \,  T^2/2$ & 6.38 fm$/c$ & 13.4 fm & 522 MeV& 432 MeV & 0.79 & 18 MeV\\
\hline
30\% & 8.9 & $0.15~\mathrm{fm}^{-1}$ & 12.85 & 0.032 $~\mathrm{fm}^{-1}$ & $4.55 \, T^2/2$ & 5.99 fm$/c$ & 12.6 fm & 491 MeV& 406 MeV & 0.82 & 15 MeV\\
\hline
40\% & 8.35 & $0.15~\mathrm{fm}^{-1}$ & 12.85 & 0.032 $~\mathrm{fm}^{-1}$ & $4.55 \, T^2/2$ & 5.59 fm$/c$ & 11.9 fm & 459 MeV& 379 MeV & 0.86 & 12 MeV\\
\hline
50\% & 7.8 & $0.15~\mathrm{fm}^{-1}$ & 12.85 & 0.032 $~\mathrm{fm}^{-1}$ & $4.55 \, T^2/2$ & 5.16 fm$/c$ & 11.1 fm & 428 MeV& 353 MeV & 0.89 & 9 MeV\\
\hline
60\% & 7.25 & $0.15~\mathrm{fm}^{-1}$ & 12.85 & 0.032 $~\mathrm{fm}^{-1}$ & $4.55 \,  T^2/2$ & 4.71 fm$/c$ & 10.3 fm & 397 MeV& 327 MeV & 0.91 & 7 MeV\\
\hline
70\% & 6.7 & $0.15~\mathrm{fm}^{-1}$ & 12.85 & 0.032 $~\mathrm{fm}^{-1}$ & $4.55 \, T^2/2$ & 4.24 fm$/c$ & 9.51 fm & 365 MeV& 301 MeV & 0.93 & 6 MeV\\
\hline
\end{tabular}
\caption{Tabulation of the parameters $\hat T_0$ and $q$  used in the construction of the Gubser solutions that we use to model collisions with differing centralities (see also Figs.~\ref{res01} and~\ref{res02}), the parameter $C_B$ that arises in our treatment of baryon stopping, and the QGP transport coefficients $\sigma$ and $\mu m$ that enter the calculation of electromagnetic-field-induced charge-dependent drift.  We also tabulate 
physical quantities characterizing the Gubser solutions:
the freezeout time $\tau_f$ at which the center of the fireball cools to $T_f=160$~MeV, the radius of the freezeout surface $R_\perp^{f}$, the initial temperature at the center $\langle T_i^{(c)}\rangle=\langle T_i(\tau=0.6,x_\perp =0)\rangle$ and the average initial temperature $\langle T_i\rangle=\langle T_i(\tau=0.6,x_\perp<7)\rangle$ inside the region $x_\perp<7$, as well as the antiproton-to-proton ratio $\bar{p}/p$ resulting from our calculation of baryon stopping, $\langle \mu_B^f \rangle$ is effective value of baryon chemical chemical potential at the freeze-out.}\label{table1}
\end{table*} 
%%%%%%%%%%%%%%%%%%%%%%%

%%%%%%%%%%%%%%%%%%%%%
%%%%%%%%%%%%%%%%%%%%%
%\newpage
\section{Results}
\label{sec:Results}
We shall present our results in this Section. Before we begin, 
in Table~\ref{table1} we enumerate all the choices of parameters that we have made in our study, including those used in the construction of the Gubser solutions that we use to model collisions with differing centralities (Sec.~\ref{sec:HydroBackground}),
the parameter that enters in our treatment of baryon stopping (Sec.~\ref{BScomp}), and those that parametrize the QGP transport properties that enter our calculation of how electromagnetic fields 
create charge-dependent directed 
flow (Secs.~\ref{EMcomp} and \ref{sec:ComputingDirectedFlow}).
The motivation for these choices was discussed in the corresponding Sections. 
The parameters $T_f$, $q$, $C_B$, $\sigma$ and $\mu m$ are the same for all centrality classes, while we have varied $\hat T_0$ so as to mimic collisions with different centralities, see Figs.~\ref{res01} and~\ref{res02}.
Specifically, we decrease $\hat T_0$ gradually in steps of $0.55$ for each $10\%$ increase in the centrality percentile. 
This serves to decrease the size and lifetime of the fireball and the resulting
freezeout surface as is appropriate, but we note that all the Gubser solutions are azimuthally symmetric.
The remaining columns of Table~\ref{table1} list various characterizations of the freezeout surfaces in the calculations that we use to model collisions with different
centralities: the freezeout time $\tau_f$, the transverse size of the freezeout surface $R_\perp^{f}$, the initial temperature at the center of the fireball  $\langle T_i(\tau=0.6,x_\perp=0)\rangle$, the average initial temperature inside the region $x_\perp<7$, denoted by $\langle T_i(\tau=0.6,x_\perp<7)\rangle$, the antiproton-to-proton ratio $\bar p/p$, and the baryon chemical chemical potential averaged over the freezeout surface, $\langle \mu_B^f \rangle$. We have discussed the compromises involved in choosing the parameters for the (azimuthally symmetric) Gubser solutions that we use to model  (azimuthally asymmetric) collisions with increasing centrality in Sections~\ref{sec:HydroBackground} and \ref{sec:Freezeout}.

%%%%%%%%%%%%%%%%%%%%%%%
\begin{figure}[t]
\centering
\includegraphics[width=1.\columnwidth]{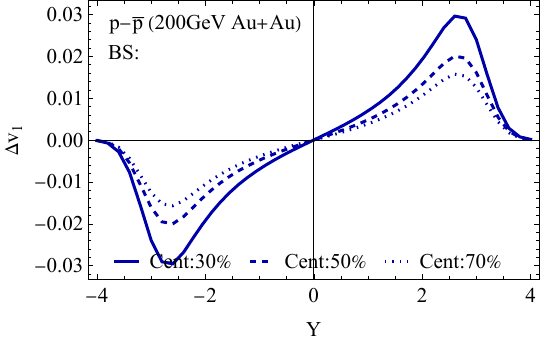}
\includegraphics[width=1.\columnwidth]{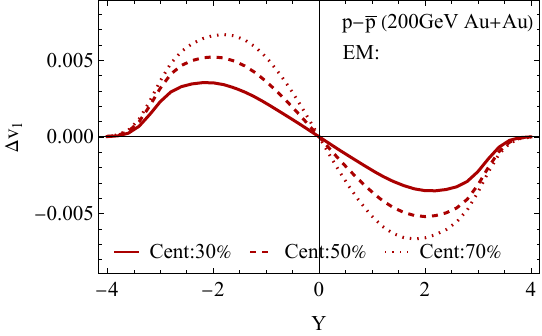}
\includegraphics[width=1.\columnwidth]{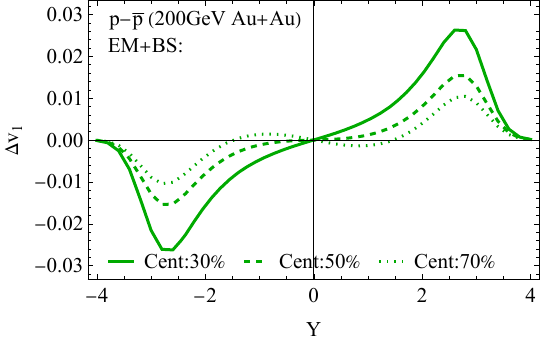}
\caption{Results from our calculations of charge-dependent directed flow $\Delta v_1(Y)\equiv v_1^p(Y)-v_1^{\bar p}(Y)$ profiles for protons/antiprotons with $p_T\in[0.4,2.0]$~GeV.
In the top panel, we include the
contribution originating from stopped baryons, but turn electromagnetic fields off.  In the middle panel, we include
the contribution originating from 
electromagnetic fields, but exclude stopped baryons. In the bottom panel, we show the results of our full calculation including both effects, EM+BS.
In each panel dotted, dashed and solid curves show our results for Au+Au collisions with
centralities  $30\%$, $50\%$ and $70\%$, respectively, and collision energy
$\sqrt{s_{NN}}=200$~GeV.  
}\label{res3}
\end{figure}
%%%%%%%%%%%%%%%%%%%%%

\begin{figure}[t]
\centering
\includegraphics[width=1.\columnwidth]{"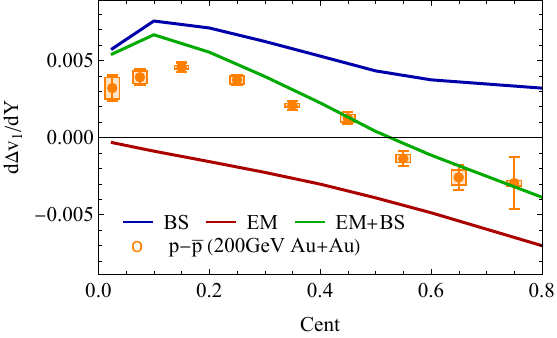"}
\caption{Centrality dependence of the 
slope of the charge-dependent proton-antiproton
directed flow at mid-rapidity, 
$d\Delta v_1/dY\equiv d v_1^p/dY-d v_1^{\bar p}/dY$
for protons/antiprotons
with $p_T\in[0.4,2.0]$~GeV. The  green curve shows the results from our full calculations including the effects of both baryon stopping and electromagnetic forces, EM+BS. 
Orange markers denote STAR measurements of $d\Delta v_1/dY$ 
for Au+Au collisions with $\sqrt{s_{NN}}=200$~GeV~\cite{STAR:2023jdd}. 
}\label{res03}
\end{figure}

\begin{figure}[t]
\centering
\includegraphics[width=1.\columnwidth]{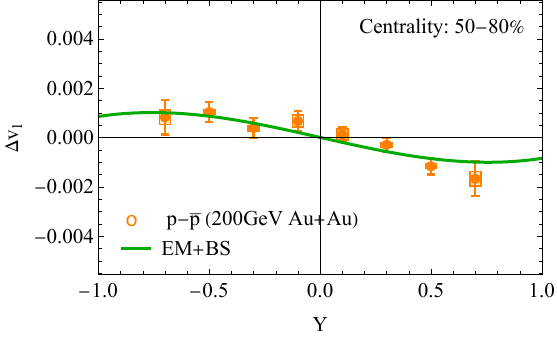}
\caption{Rapidity dependence of the charge-dependent proton-antiproton directed flow,  $\Delta v_1 \equiv v_1(p)-v_1(\bar{p})$, 
for protons/antiprotons with
$p_T\in[0.4,2.0]$~GeV
in Au+Au collisions with $\sqrt{s_{NN}}=200$~GeV
within a broad range of centralities, 
$50\%-80\%$, where effects originating from electromagnetic forces dominate. 
The  green curve denotes the
results of our full calculation including effects originating from electromagnetic forces and baryon stopping, EM+BS,
which can be compared to experimental data. Orange markers denote STAR measurements of $\Delta v_1(Y)$ 
for Au+Au collisions with $\sqrt{s_{NN}}=200$~GeV~\cite{STAR:2023jdd}. 
}\label{res002}
\end{figure}

We present the main results of this study in Figs.~\ref{res3}, \ref{res03} and \ref{res002}.
In Fig.~\ref{res3}, we show
the results of our model calculations of
the rapidity dependence of the proton--antiproton directed-flow splitting, $\Delta v_1(Y)\equiv v_1(p)-v_1(\bar p)$, for Au+Au collisions at $\sqrt{s_{NN}}=200$~GeV. As in the STAR analysis of experimental data~\cite{STAR:2023jdd}, we only include protons and antiprotons with $0.4~{\rm GeV}\leq p_T \leq 2.0~{\rm GeV}$.  The top, middle, and bottom panels correspond, respectively, to the baryon-stopping contribution alone, the electromagnetic contribution alone, and the results from our full calculation including the combined contribution originating from both mechanisms acting in concert. Throughout this Section, we use the same color code to distinguish the different cases: blue for baryon stopping only, red for electromagnetic effects only, and green for the combined result. In each panel, the solid, dashed, and dotted curves denote the results of our calculations for heavy ion collisions in the $30\%$, $50\%$, and $70\%$ centrality classes, respectively.

The top panel of Fig.~\ref{res3}
shows that baryon stopping generates a rapidity-odd $\Delta v_1$ with a positive rapidity-slope around $Y=0$,
as we described in our discussion of Fig.~\ref{coll_geo}.  The magnitude of $\Delta v_1$
decreases as one moves from mid-central to more peripheral collisions
reflecting the decreasing 
importance of the rapidity-odd and reflection-odd component of the stopped proton distribution in peripheral collisions.
The rapidity-odd and reflection-odd component of the stopped baryon distribution vanishes by symmetry for head-on collisions with impact parameter $b=0$, and also vanishes for grazing collisions with $b=2R$, $R$ being the nuclear radius.  It is maximal for collisions with $b\sim R$, which per Fig.~\ref{cofb} corresponds to collisions with 10-20\% centrality.  Given this, it is no surprise to see in the top panel of Fig.~\ref{res3}
that the charge-dependent rapidity-odd directed flow $\Delta v_1$ originating from
stopped protons decreases in magnitude as one goes from collisions with 30\% centrality to 50\% to 70\%.

The middle panel of Fig.~\ref{res3} shows 
that the forces originating from electromagnetic fields yield
a $\Delta v_1$ that is also rapidity-odd, but in other respects has the opposite behavior to that originating from stopped protons. In the middle panel, $\Delta v_1(Y)$ 
has a negative rapidity-slope at midrapidity, as we described in our discussion of Fig.~\ref{coll_geo}.
This confirms that the Faraday and Coulomb forces originating from the time-derivative of the
magnetic field and the electric field dominate over the Lorentz force originating from longitudinal fluid motion in the transverse 
magnetic field, as found in previous studies~\cite{Gursoy:2014aka,Gursoy:2018yai}.  The magnitude of $\Delta v_1(Y)$
in the middle panel increases
toward more peripheral collisions,
where there are more spectators and so the spectator-induced magnetic field
is larger.

The difference between the baryon-stopping and electromagnetic contributions to the charge-dependent directed flow is not limited to the opposite signs of their midrapidity slopes: their rapidity profiles are also visibly different.  The baryon-stopping contribution extends further toward forward and backward rapidities, with its magnitude peaking around $|Y|\simeq 2.6$--$2.8$, whereas the magnitude of the electromagnetic contribution peaks closer to midrapidity, around $|Y|\simeq 1.8$--$2.0$, and then decreases toward larger $|Y|$.  The former behavior reflects the rapidity-distribution of the stopped baryons, as it is their rapidity-dependent transverse asymmetry that is converted into charge-dependent directed flow by radial expansion.  
The decrease in the electromagnetic contribution at the largest rapidities was described previously in Ref.~\cite{Gursoy:2018yai}: it arises from a near-cancellation between the Coulomb+Faraday and Lorentz contributions (which have opposite sign, see Fig.~\ref{coll_geo}) at large spacetime-rapidity $\eta_s$.
These difference between the rapidity profiles
of the electromagnetic and baryon-stopping contributions to the charge-dependent directed flow provides an additional experimental handle via which to discern each of these contributions. This motivates extending the
measurements of $\Delta v_1(Y)$ 
to the broadest possible rapidity range.

The bottom panel of Fig.~\ref{res3} displays the combined result of the two mechanisms acting in concert, and serves to illustrate a central message of this study.  Our calculations confirm the intuition from Ref.~\cite{STAR:2023jdd}:
for heavy ion collisions in the
$30\%$ centrality class, the contribution to $\Delta v_1(Y)$ coming from stopped protons dominates and the slope $d \Delta v_1/dY$ at $Y=0$
remains positive;
for peripheral collisions in the 70\% centrality class, the
contribution to $\Delta v_1(Y)$ coming from electromagnetic forces dominates
for $|Y|\lesssim 1$
and  $d \Delta v_1/dY$ at $Y=0$ is negative,
as in Refs.~\cite{Gursoy:2014aka,Gursoy:2018yai}; and, around
the $50\%$ centrality class, the two effects nearly compensate each other,
and the slope at $Y=0$ is small near where it changes sign.
Thus, Fig.~\ref{res3} provides a direct visualization of how the competition between baryon stopping and electromagnetic forces produces a centrality-dependent sign change of $\Delta v_1$, as seen
in experimental 
data~\cite{STAR:2023jdd}.
Our results yield the qualitative prediction that if it is possible in future to extend the measurements of $\Delta v_1(Y)$ to $|Y|\gtrsim 2$, in this high-rapidity regime the contribution to this observable originating from baryon stopping should dominate, meaning that this observable should not change sign with increasing centrality.

%%%%%%%%%%%%%%%%%%%%%
%%%%%%%%%%%%%%%%%%%%%%%

In Fig.~\ref{res03}, we show how
the 
slope of the proton--antiproton charge-dependent directed flow  $d\Delta v_1/dY$
at midrapidity changes with collision centrality in
Au+Au collisions at $\sqrt{s_{NN}}=200$~GeV in our calculations, and compare to the experimental data from the STAR measurement of Ref.~\cite{STAR:2023jdd}. 
We use the same color convention as in Fig.~\ref{res3}: the blue curve shows the contribution originating from stopped protons alone, the red curve shows the contribution originating from electromagnetic forces alone, and the green curve shows the full result of our model calculation with both mechanisms acting in concert, which can be compared to the STAR data, shown in orange.

The separate contributions 
originating from stopped protons (blue curve) and electromagnetic forces (red curve) in Fig.~\ref{res03}
have the expected
opposite signs over the full centrality range. 
The combined result shown in the green curve
reproduces the qualitative trend seen in the STAR data and, in particular, reproduces the observed sign change from positive to negative slope as the collision becomes more peripheral. 
Our calculations are in good qualitative agreement with the STAR data across the full range of centrality, with
the agreement being especially good in the midcentral region, where the sign change occurs.
Fig.~\ref{res03} highlights key findings of our study:
neither effects originating from electromagnetic forces nor effects 
originating from stopped protons 
alone yield even a qualitative description of the centrality dependence of $d\Delta v_1/dY$ seen in experimental data,
while the competition between them that arises when both mechanisms act in concert 
naturally reproduces the sign reversal 
with increasing centrality seen in experimental data, and in fact yields a good qualitative description of this data across the full range of centrality.
We note that the sign of the slight discrepancy between the results of our calculation and the STAR data for the most central collisions is easily understood: event-by-event fluctuations in the shape of the collision-overlap zone and hence in the number of spectators that are not included in our calculations would result in electromagnetic fields and consequent forces in the most central collisions that are not as small in magnitude as in our model. 

%%%%%%%%%%%%%%%%%%%%%%%

In 
Fig.~\ref{res002}, 
we
show the rapidity dependence of the proton--antiproton directed-flow splitting, $\Delta v_1(Y)=v_1(p)-v_1(\bar p)$, for 
Au+Au collisions with $\sqrt{s_{NN}}=200$~GeV 
in the $50$--$80\%$ centrality interval.
The slope of the green curve
depicting the results of our full EM+BS calculation is a suitably weighted average of the 
results between 50\% and 80\% centrality from the green curve in Fig.~\ref{res03},
combined using the Glauber weights 
(\ref{eq:Cofb}) introduced in Sec.~\ref{gb-section}.
The orange points with error bars denote the experimental data from the STAR measurement of Ref.~\cite{STAR:2023jdd}.
The 50\%-80\% centrality range corresponds to the range of centralities in which the contributions originating from electromagnetic forces are larger than the contributions originating from stopped baryons.  The excellent qualitative agreement between our calculations and the STAR data in 
Fig.~\ref{res002}
across the whole range of rapidity shown
tests the model beyond the mid-rapidity-slope comparison shown in Fig.~\ref{res03}, and shows that it
provides a fully successful qualitative description of the experimental data
where the electromagnetic contribution is the most important.
\begin{figure}[t]
\centering
\includegraphics[width=1.\columnwidth]{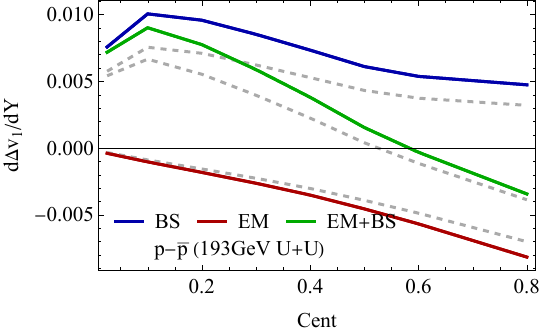}
\caption{
Centrality dependence of the 
slope of the charge-dependent proton-antiproton
directed flow at mid-rapidity, 
$d\Delta v_1/dY\equiv d v_1^p/dY-d v_1^{\bar p}/dY$
for protons/antiprotons
with $p_T\in[0.4,2.0]$~GeV
in U+U collisions with 
$\sqrt{s_{NN}}=193$~GeV.
The  green curve shows the results from our full calculations including the effects of both baryon stopping and electromagnetic forces, EM+BS. 
As in previous Figures, the blue and red curves include only baryon stopping and only electromagnetic effects. They dashed grey curves show the corresponding results from our calculations for Au+Au collisions, from Fig.~\ref{res03}. 
}\label{res001}
\end{figure}

In Fig.~\ref{res001},
we show results from our calculations that are analogous to those in Fig.~\ref{res3}, here for U+U collisions
 at $\sqrt{s_{NN}}=193$~GeV.
The green curve shows
the centrality dependence of the midrapidity slope of the proton--antiproton directed-flow splitting, $d\Delta v_1/dY$ from our calculations
that include both baryon stopping
and electromagnetic forces.
The 
same color convention is used as in  previous figures: the blue and red curves show the contributions from baryon stopping alone and from electromagnetic forces alone, respectively. For comparison, the
grey dashed curves 
show the corresponding results for Au+Au collisions at $\sqrt{s_{NN}}=200$~GeV from Fig~\ref{res3}.

Fig.~\ref{res001} is motivated by the STAR measurement of 
charge-dependent directed flow in U+U collisions. Preliminary data from this measurment was reported at the Strangeness in Quark Matter (SQM) 2024 conference~\cite{Taseer:2024sho}. U+U collisions provide an interesting test of our model because uranium has a larger charge and atomic number than gold, leading, at comparable centrality, to different numbers of spectators and participants. 
The former directly affects the strength of the spectator-induced electromagnetic fields, while the latter is related to the amount of baryon stopping. Therefore, comparing U+U with Au+Au is a good way to further test whether 
our model captures the qualitative 
behavior of the charge-dependent directed flow that results from the two competing mechanisms acting in concert.
Although we have not plotted STAR 
data in Fig.~\ref{res001} because this measurement has not yet been submitted for publication, it is easy to check that our results (the green curve in Fig.~\ref{res001}) provide a very good qualitative description of the preliminary STAR data reported in Ref.~\cite{Taseer:2024sho}, including the shift in $d\Delta v_1/dY$ as a function of centrality in U+U collisions relative to that measured in Au+Au collisions with the centrality at which the $d\Delta v_1/dY$ changes sign shifting slightly towards more peripheral collisions.  
This indicates that
the increased fluctuations in the number of spectators and participants in mid-rapidity collisions on account of the deformed shape of uranium nuclei (which is not included in our model) does not affect this observable so dramatically as to spoil the qualitative agreement between our model and experimental data.

We leave modeling the charge-dependent directed flow of other hadronic species to future work. $\Delta v_1$ for charged pions and kaons should receive an electromagnetic contribution that we could estimate using the
methods of this paper, but the calculation of their $\Delta v_1$ is rendered less direct and more challenging by the fact that
a large fraction of these light mesons originate from the decays of resonances, which will dilute electromagnetic effects and, since many originate from 
baryon resonances, introduce indirect effects of baryon stopping.
The $\Delta v_1$ for kaons
will also receive a direct contribution from baryon stopping since the $u$ quark in a $K^+$ can originate from stopped protons 
whereas the $\bar u$ antiquark in a 
$K^-$ cannot, which necessitates a more sophisticated approach to modeling the mechanism and effects of baryon stopping,  rather than just following the contribution from stopped baryons to $\mu_B$ as we have done.  
Similar considerations would apply to modeling $\Delta v_1$ for $D$ mesons or $\Lambda_c$ baryons, all of which we leave to future work.

%%%%%%%%%%%%%%%%%%%%%
\section{Concluding Remarks and a Look Ahead} 
\label{sec:Conclusion}
%%%%%%%%%%%%%%%%%%%%%
%%%%%%%%%%%%%%%%%%%%%
In this work, we have developed a semi-analytic approach to analyze the charge-dependent directed flow in heavy-ion collisions by combining the effects of spectator-induced electromagnetic fields with baryon stopping within a single unified framework. We have modeled the hydrodynamic background using Gubser flow, computed the electromagnetic field from the charged spectators in a conducting medium, and implemented the charge-dependent directed flow caused by electromagnetic forces
through a force-balance drift correction. We have incorporated baryon stopping starting from a Glauber model stopped-baryon profile, which generates a spacetime-dependent baryon chemical potential on the freezeout surface that modifies the relative emission of protons and antiprotons. The distribution of stopped protons is odd in rapidity and odd under reflection in the impact parameter axis; the boost imparted to this distribution by the radial flow (described in our model by the Gubser flow background hydrodynamic solution)
means that this also results in charge-dependent directed flow.

Applying this framework to Au+Au collisions at $\sqrt{s_{NN}}=200$ GeV, we found that the two mechanisms contribute with opposite signs,
and with different dependence on the centrality of the collision,
to the difference between the directed flow of protons and antiprotons $\Delta v_1$. 
Baryon stopping yields a positive contribution to the slope
$d \Delta v_1/dY$ at midrapidity which dominates in head-on and midcentral
collisions while the electromagnetic forces yield a negative contribution to
$d \Delta v_1/dY$ at midrapidity
that becomes stronger toward more peripheral collisions.
This means that these two effects in concert naturally result in a change in the sign of
$d\Delta v_1/dY$ at midrapidity as a function of centrality, in qualitative agreement with experimental measurements from STAR. 
The model also gives a good description of the measured $\Delta v_1(Y)$ profile in the $50$--$80\%$ centrality interval and captures the expected system-size trend when extended to U+U collisions. Our results further show that the baryon-stopping and electromagnetic contributions differ not only in the signs of their midrapidity slopes but also in their rapidity dependence, with the contribution from baryon stopping extending out to larger $|Y|$ than that caused by electromagnetic forces, 
and dominating at $|Y|\gtrsim 2$ for all collision centralities. 
This provides an additional experimental handle for disentangling the two effects and motivates measuring $\Delta v_1(Y)$ over the widest possible rapidity range.

At the same time, the framework that we have developed and presented here has important limitations --- each of which points the way forward toward important future work. We highlight three examples:
\begin{itemize}
\item
Because doing so makes a semi-analytical approach possible, we have built the present model on hydrodynamic backgrounds given by Gubser's solution, using different choices of the parameter $\hat T_0$ to model the hydrodynamic droplets of QGP produced in heavy ion collisions with different centralities.
With any choice of its parameters, Gubser's solution is azimuthally symmetric whereas a heavy ion collision with nonzero impact parameter is not. This motivates the implementation of our unified calculation of the contribution to charge-dependent directed flow sourced by electromagnetic forces and baryon stopping in backgrounds obtained from numerical relativistic viscous hydrodynamic calculations that provide a realistic description of heavy ion collisions with nonzero impact parameter.
First steps in this direction have already been taken in Ref.~\cite{Parida:2025ddt}, although these authors take a different approach to implementing baryon stopping in which the rapidity dependence of the distribution of stopped baryons does not take the empirical form~\eqref{eq:feta_BS}~\cite{KHARZEEV1996238,KHARZEEV2008227,ALICE:2013yba,Frenklakh:2024mgu}
meaning that its ``tilt'' (which determines how much it contributes to the charge-dependent directed flow) is not related to, and therefore not constrained by, the net-proton yield.

\item
Although the use of an empirically constrained stopped baryon distribution is a strength of our model, the treatment of the consequences of the stopped baryon distribution that we have implemented in the framework that we have developed in this work can also be further improved. In particular, since we have chosen to assess the impact of the rapidity- and reflection-odd component of the stopped baryon distribution only via tracing
the resulting spacetime-dependent baryon chemical potential across the freezeout hypersurface, not by tracing the evolution of the distribution of stopped quarks with each flavor and following the dynamics of all conserved charges, we can calculate the difference between the proton and antiproton 
directed flow but we do not have access to the charge-dependent directed flow of mesons.  
This motivates a further extension of our study, which would at the same time open the door to improved treatments of hadronization and freezeout.  

\item
Last, we highlight once again that it is an oversimplification to treat the electrical conductivity of the QGP as a constant, as we have done. There is certainly strong motivation to pursue magnetohydrodynamic calculations that incorporate a temperature-dependent conductivity.
\end{itemize}
Neither improving the treatment of the consequences of baryon stopping so as to follow the evolution of individual flavors of quarks and all conserved charges nor pursuing magnetohydrodynamic calculations with an electrical conductivity that varies in space and time is motivated in a model where we use the Gubser solution as the hydrodynamic background.  But, once our approach has been implemented in realistic hydrodynamic backgrounds both of these other advances will become necessary steps on the road to making quantitative predictions for experimental measurements of the charge-dependent directed flow in heavy ion collisions.
Subsequently, going further down the road to precision will require event-by-event hydrodynamic simulations including fluctuations in the electromagnetic fields and baryon stopping.

Despite its simplifications, the model framework that we have introduced here is
a useful tool because at the same time that it provides a unified framework with which to assess the combined effects of baryon stopping and electromagnetic forces on the charge-dependent directed flow it allows us to assess their competition in a controlled way.
Each mechanism generates a contribution to the difference between the proton and antiproton directed flow with an opposite sign and with very different dependences on rapidity and collision centrality. Although the present modeling framework does not provide quantitative predictions, it provides a clear and physically transparent explanation of the qualitative features seen in  the centrality- and rapidity-dependence of
recent experimental measurements of the difference between the proton and antiproton directed flow.

\begin{acknowledgments}
The early phase of this research was done in collaboration with Umut G\"ursoy; we miss him, his wisdom, and his physics intuition, enormously.
We are grateful to Andrea Dubla, Govert Nijs, Ilya Selyuzhenkov, Wilke van der Schee, Chun Shen, Aihong Tang and Raul Wolters for helpful conversations.
KR is grateful to the CERN Theory Department for hospitality as this work was completed.
Research supported in part by the U.S.~Department of Energy, Office of Science, Office of Nuclear Physics under grant Contract Number DE-SC0011090. The work of DK was supported in part by the U.S. Department of Energy, Office of Science, Office of Nuclear Physics under Contracts No. DE-SC0026415 and No. DE-FG02-88ER40388. TD and RS have been supported by the Netherlands Organization for Scientific Research (NWO) with the grants VICI–BN.000665.1 and ENWXL–BN.000704.1. 
\end{acknowledgments}

\bibliography{refs.bib}

@article{Astrakhantsev:2019zkr,
    author = "Astrakhantsev, Nikita and Braguta, V. V. and D'Elia, Massimo and Kotov, A. Yu. and Nikolaev, A. A. and Sanfilippo, Francesco",
    title = "{Lattice study of the electromagnetic conductivity of the quark-gluon plasma in an external magnetic field}",
    eprint = "1910.08516",
    archivePrefix = "arXiv",
    primaryClass = "hep-lat",
    doi = "10.1103/PhysRevD.102.054516",
    journal = "Phys. Rev. D",
    volume = "102",
    number = "5",
    pages = "054516",
    year = "2020"
}

@article{Aarts:2020dda,
    author = "Aarts, Gert and Nikolaev, Aleksandr",
    title = "{Electrical conductivity of the quark-gluon plasma: perspective from lattice QCD}",
    eprint = "2008.12326",
    archivePrefix = "arXiv",
    primaryClass = "hep-lat",
    doi = "10.1140/epja/s10050-021-00436-5",
    journal = "Eur. Phys. J. A",
    volume = "57",
    number = "4",
    pages = "118",
    year = "2021"
}

@article{Amato:2013naa,
    author = "Amato, Alessandro and Aarts, Gert and Allton, Chris and Giudice, Pietro and Hands, Simon and Skullerud, Jon-Ivar",
    title = "{Electrical conductivity of the quark-gluon plasma across the deconfinement transition}",
    eprint = "1307.6763",
    archivePrefix = "arXiv",
    primaryClass = "hep-lat",
    doi = "10.1103/PhysRevLett.111.172001",
    journal = "Phys. Rev. Lett.",
    volume = "111",
    number = "17",
    pages = "172001",
    year = "2013"
}

@article{Brandt:2012jc,
    author = "Brandt, Bastian B. and Francis, Anthony and Meyer, Harvey B. and Wittig, Hartmut",
    title = "{Thermal Correlators in the {\textbackslash}rho{\textbackslash} channel of two-flavor QCD}",
    eprint = "1212.4200",
    archivePrefix = "arXiv",
    primaryClass = "hep-lat",
    doi = "10.1007/JHEP03(2013)100",
    journal = "JHEP",
    volume = "03",
    pages = "100",
    year = "2013"
}

@article{Francis:2011bt,
    author = "Francis, A. and Kaczmarek, O.",
    editor = "Faessler, Amand and Rodin, Vadim",
    title = "{On the temperature dependence of the electrical conductivity in hot quenched lattice QCD}",
    eprint = "1112.4802",
    archivePrefix = "arXiv",
    primaryClass = "hep-lat",
    doi = "10.1016/j.ppnp.2011.12.020",
    journal = "Prog. Part. Nucl. Phys.",
    volume = "67",
    pages = "212--217",
    year = "2012"
}

@article{Burnier:2011bf,
    author = "Burnier, Yannis and Kharzeev, Dmitri E. and Liao, Jinfeng and Yee, Ho-Ung",
    title = "{Chiral magnetic wave at finite baryon density and the electric quadrupole moment of quark-gluon plasma in heavy ion collisions}",
    eprint = "1103.1307",
    archivePrefix = "arXiv",
    primaryClass = "hep-ph",
    doi = "10.1103/PhysRevLett.107.052303",
    journal = "Phys. Rev. Lett.",
    volume = "107",
    pages = "052303",
    year = "2011"
}

@article{Voloshin:2010ut,
    author = "Voloshin, Sergei A.",
    title = "{Testing the Chiral Magnetic Effect with Central U+U collisions}",
    eprint = "1006.1020",
    archivePrefix = "arXiv",
    primaryClass = "nucl-th",
    doi = "10.1103/PhysRevLett.105.172301",
    journal = "Phys. Rev. Lett.",
    volume = "105",
    pages = "172301",
    year = "2010"
}

@article{Schlichting:2010qia,
    author = "Schlichting, Soren and Pratt, Scott",
    title = "{Charge conservation at energies available at the BNL Relativistic Heavy Ion Collider and contributions to local parity violation observables}",
    eprint = "1009.4283",
    archivePrefix = "arXiv",
    primaryClass = "nucl-th",
    doi = "10.1103/PhysRevC.83.014913",
    journal = "Phys. Rev. C",
    volume = "83",
    pages = "014913",
    year = "2011"
}

@article{Kharzeev:2022hqz,
    author = "Kharzeev, Dmitri E. and Liao, Jinfeng and Shi, Shuzhe",
    title = "{Implications of the isobar-run results for the chiral magnetic effect in heavy-ion collisions}",
    eprint = "2205.00120",
    archivePrefix = "arXiv",
    primaryClass = "nucl-th",
    doi = "10.1103/PhysRevC.106.L051903",
    journal = "Phys. Rev. C",
    volume = "106",
    number = "5",
    pages = "L051903",
    year = "2022"
}

@article{STAR:2021mii,
    author = "Abdallah, Mohamed and others",
    collaboration = "STAR",
    title = "{Search for the chiral magnetic effect with isobar collisions at $\sqrt {s_{NN}}$=200 GeV by the STAR Collaboration at the BNL Relativistic Heavy Ion Collider}",
    eprint = "2109.00131",
    archivePrefix = "arXiv",
    primaryClass = "nucl-ex",
    doi = "10.1103/PhysRevC.105.014901",
    journal = "Phys. Rev. C",
    volume = "105",
    number = "1",
    pages = "014901",
    year = "2022"
}

@article{Bzdak:2012ia,
    author = "Bzdak, Adam and Koch, Volker and Liao, Jinfeng",
    title = "{Charge-Dependent Correlations in Relativistic Heavy Ion Collisions and the Chiral Magnetic Effect}",
    eprint = "1207.7327",
    archivePrefix = "arXiv",
    primaryClass = "nucl-th",
    reportNumber = "RBRC-964",
    doi = "10.1007/978-3-642-37305-3_19",
    journal = "Lect. Notes Phys.",
    volume = "871",
    pages = "503--536",
    year = "2013"
}

@article{Brandt:2015aqk,
    author = {Brandt, Bastian B. and Francis, Anthony and J{\"a}ger, Benjamin and Meyer, Harvey B.},
    title = "{Charge transport and vector meson dissociation across the thermal phase transition in lattice QCD with two light quark flavors}",
    eprint = "1512.07249",
    archivePrefix = "arXiv",
    primaryClass = "hep-lat",
    reportNumber = "MITP-15-019",
    doi = "10.1103/PhysRevD.93.054510",
    journal = "Phys. Rev. D",
    volume = "93",
    number = "5",
    pages = "054510",
    year = "2016"
}

@article{Ding:2016hua,
    author = "Ding, Heng-Tong and Kaczmarek, Olaf and Meyer, Florian",
    title = "{Thermal dilepton rates and electrical conductivity of the QGP from the lattice}",
    eprint = "1604.06712",
    archivePrefix = "arXiv",
    primaryClass = "hep-lat",
    reportNumber = "BI-TP-2016-03",
    doi = "10.1103/PhysRevD.94.034504",
    journal = "Phys. Rev. D",
    volume = "94",
    number = "3",
    pages = "034504",
    year = "2016"
}

@article{Aarts:2014nba,
    author = "Aarts, Gert and Allton, Chris and Amato, Alessandro and Giudice, Pietro and Hands, Simon and Skullerud, Jon-Ivar",
    title = "{Electrical conductivity and charge diffusion in thermal QCD from the lattice}",
    eprint = "1412.6411",
    archivePrefix = "arXiv",
    primaryClass = "hep-lat",
    reportNumber = "HIP-2014-34-TH, INT-PUB-14-060, MS-TP-14-40",
    doi = "10.1007/JHEP02(2015)186",
    journal = "JHEP",
    volume = "02",
    pages = "186",
    year = "2015"
}

@article{Ding:2010ga,
    author = "Ding, H. -T. and Francis, A. and Kaczmarek, O. and Karsch, F. and Laermann, E. and Soeldner, W.",
    title = "{Thermal dilepton rate and electrical conductivity: An analysis of vector current correlation functions in quenched lattice QCD}",
    eprint = "1012.4963",
    archivePrefix = "arXiv",
    primaryClass = "hep-lat",
    reportNumber = "BI-TP-2010-46",
    doi = "10.1103/PhysRevD.83.034504",
    journal = "Phys. Rev. D",
    volume = "83",
    pages = "034504",
    year = "2011"
}

@article{Nijs:2020ors,
    author = {Nijs, Govert and van der Schee, Wilke and G{\"u}rsoy, Umut and Snellings, Raimond},
    title = "{Transverse Momentum Differential Global Analysis of Heavy-Ion Collisions}",
    eprint = "2010.15130",
    archivePrefix = "arXiv",
    primaryClass = "nucl-th",
    reportNumber = "CERN-TH-2020-174, MIT-CTP/5250",
    doi = "10.1103/PhysRevLett.126.202301",
    journal = "Phys. Rev. Lett.",
    volume = "126",
    number = "20",
    pages = "202301",
    year = "2021"
}

@article{JETSCAPE:2020mzn,
    author = "Everett, D. and others",
    collaboration = "JETSCAPE",
    title = "{Multisystem Bayesian constraints on the transport coefficients of QCD matter}",
    eprint = "2011.01430",
    archivePrefix = "arXiv",
    primaryClass = "hep-ph",
    doi = "10.1103/PhysRevC.103.054904",
    journal = "Phys. Rev. C",
    volume = "103",
    number = "5",
    pages = "054904",
    year = "2021"
}

@article{Parkkila:2021tqq,
    author = "Parkkila, J. E. and Onnerstad, A. and Kim, D. J.",
    title = "{Bayesian estimation of the specific shear and bulk viscosity of the quark-gluon plasma with additional flow harmonic observables}",
    eprint = "2106.05019",
    archivePrefix = "arXiv",
    primaryClass = "hep-ph",
    doi = "10.1103/PhysRevC.104.054904",
    journal = "Phys. Rev. C",
    volume = "104",
    number = "5",
    pages = "054904",
    year = "2021"
}

@article{Snellings:1999bt,
    author = "Snellings, R. J. M. and Sorge, H. and Voloshin, S. A. and Wang, F. Q. and Xu, N.",
    title = "{Novel rapidity dependence of directed flow in high-energy heavy ion collisions}",
    eprint = "nucl-ex/9908001",
    archivePrefix = "arXiv",
    doi = "10.1103/PhysRevLett.84.2803",
    journal = "Phys. Rev. Lett.",
    volume = "84",
    pages = "2803--2805",
    year = "2000"
}

@article{Benoit:2025amn,
    author = "Benoit, Nicholas J. and Miyoshi, Takahiro and Nonaka, Chiho and Takahashi, Hiroyuki R.",
    title = "{Investigating effects of the electrical conductivity of QCD matter on charge-dependent directed flow}",
    eprint = "2502.04611",
    archivePrefix = "arXiv",
    primaryClass = "nucl-th",
    reportNumber = "HUPD-2502",
    doi = "10.1103/8trh-rd6d",
    journal = "Phys. Rev. C",
    volume = "112",
    number = "2",
    pages = "024911",
    year = "2025"
}

@article{Parkkila:2021yha,
    author = "Parkkila, J. E. and Onnerstad, A. and Taghavi, S. F. and Mordasini, C. and Bilandzic, A. and Virta, M. and Kim, D. J.",
    title = "{New constraints for QCD matter from improved Bayesian parameter estimation in heavy-ion collisions at LHC}",
    eprint = "2111.08145",
    archivePrefix = "arXiv",
    primaryClass = "hep-ph",
    doi = "10.1016/j.physletb.2022.137485",
    journal = "Phys. Lett. B",
    volume = "835",
    pages = "137485",
    year = "2022"
}

@article{Miller:2007ri,
    author = "Miller, Michael L. and Reygers, Klaus and Sanders, Stephen J. and Steinberg, Peter",
    title = "{Glauber modeling in high energy nuclear collisions}",
    eprint = "nucl-ex/0701025",
    archivePrefix = "arXiv",
    doi = "10.1146/annurev.nucl.57.090506.123020",
    journal = "Ann. Rev. Nucl. Part. Sci.",
    volume = "57",
    pages = "205--243",
    year = "2007"
}

@article{HotQCD:2014kol,
    author = "Bazavov, A. and others",
    collaboration = "HotQCD",
    title = "{Equation of state in (2+1)-flavor QCD}",
    eprint = "1407.6387",
    archivePrefix = "arXiv",
    primaryClass = "hep-lat",
    reportNumber = "BNL-105928-2014-JA",
    doi = "10.1103/PhysRevD.90.094503",
    journal = "Phys. Rev. D",
    volume = "90",
    pages = "094503",
    year = "2014"
}

@article{Borsanyi:2013bia,
    author = "Borsanyi, Szabolcs and Fodor, Zoltan and Hoelbling, Christian and Katz, Sandor D. and Krieg, Stefan and Szabo, Kalman K.",
    title = "{Full result for the QCD equation of state with 2+1 flavors}",
    eprint = "1309.5258",
    archivePrefix = "arXiv",
    primaryClass = "hep-lat",
    doi = "10.1016/j.physletb.2014.01.007",
    journal = "Phys. Lett. B",
    volume = "730",
    pages = "99--104",
    year = "2014"
}

@article{PHENIX:2003iij,
    author = "Adler, S. S. and others",
    collaboration = "PHENIX",
    title = "{Identified charged particle spectra and yields in Au+Au collisions at S(NN)**1/2 = 200-GeV}",
    eprint = "nucl-ex/0307022",
    archivePrefix = "arXiv",
    doi = "10.1103/PhysRevC.69.034909",
    journal = "Phys. Rev. C",
    volume = "69",
    pages = "034909",
    year = "2004"
}

@misc{Parida:2025ddt,
    author = "Parida, Tribhuban and Chatterjee, Sandeep and Singha, Subhash",
    title = "{Charge dependent directed flow splitting from baryon inhomogeneity and electromagnetic field}",
    eprint = "2503.04660",
    archivePrefix = "arXiv",
    primaryClass = "nucl-th",
}

@misc{Beraudo:2025nvq,
    author = "Beraudo, Andrea and Du Plessis, Jean F. and Pablos, Daniel and Rajagopal, Krishna",
    title = "{Heavy Quark Energy Loss in the Hybrid Model}",
    eprint = "2510.24847",
    archivePrefix = "arXiv",
    primaryClass = "hep-ph",
    reportNumber = "MIT-CTP/5950",
    month = "10",
    year = "2025"
}

@article{Herzog:2006gh,
    author = "Herzog, C. P. and Karch, A. and Kovtun, P. and Kozcaz, C. and Yaffe, L. G.",
    title = "{Energy loss of a heavy quark moving through N=4 supersymmetric Yang-Mills plasma}",
    eprint = "hep-th/0605158",
    archivePrefix = "arXiv",
    reportNumber = "NSF-KITP-06-36",
    doi = "10.1088/1126-6708/2006/07/013",
    journal = "JHEP",
    volume = "07",
    pages = "013",
    year = "2006"
}

@article{Gubser:2006bz,
    author = "Gubser, Steven S.",
    title = "{Drag force in AdS/CFT}",
    eprint = "hep-th/0605182",
    archivePrefix = "arXiv",
    reportNumber = "PUPT-2198",
    doi = "10.1103/PhysRevD.74.126005",
    journal = "Phys. Rev. D",
    volume = "74",
    pages = "126005",
    year = "2006"
}

@article{Gubser:2010ze,
    author = "Gubser, Steven S.",
    title = "{Symmetry constraints on generalizations of Bjorken flow}",
    eprint = "1006.0006",
    archivePrefix = "arXiv",
    primaryClass = "hep-th",
    reportNumber = "PUPT-2340",
    doi = "10.1103/PhysRevD.82.085027",
    journal = "Phys. Rev. D",
    volume = "82",
    pages = "085027",
    year = "2010"
}

@article{Gursoy:2018yai,
    author = {G{\"u}rsoy, Umut and Kharzeev, Dmitri and Marcus, Eric and Rajagopal, Krishna and Shen, Chun},
    title = "{Charge-dependent Flow Induced by Magnetic and Electric Fields in Heavy Ion Collisions}",
    eprint = "1806.05288",
    archivePrefix = "arXiv",
    primaryClass = "hep-ph",
    doi = "10.1103/PhysRevC.98.055201",
    journal = "Phys. Rev. C",
    volume = "98",
    number = "5",
    pages = "055201",
    year = "2018"
}

@article{ALICE:2019sgg,
    author = "Acharya, Shreyasi and others",
    collaboration = "ALICE",
    title = "{Probing the effects of strong electromagnetic fields with charge-dependent directed flow in Pb-Pb collisions at the LHC}",
    eprint = "1910.14406",
    archivePrefix = "arXiv",
    primaryClass = "nucl-ex",
    reportNumber = "CERN-EP-2019-241",
    doi = "10.1103/PhysRevLett.125.022301",
    journal = "Phys. Rev. Lett.",
    volume = "125",
    number = "2",
    pages = "022301",
    year = "2020"
}

@article{Taseer:2024sho,
    author = "Taseer, Muhammad Farhan",
    title = "{Measurement of charge-dependent directed flow in STAR Beam Energy Scan (BES-II) Au+Au and U+U collisions}",
    eprint = "2412.18326",
    archivePrefix = "arXiv",
    primaryClass = "nucl-ex",
    doi = "10.1051/epjconf/202531606008",
    journal = "EPJ Web Conf.",
    volume = "316",
    pages = "06008",
    year = "2025"
}

@article{STAR:2023jdd,
    author = "Abdulhamid, M. I. and others",
    collaboration = "STAR",
    title = "{Observation of the electromagnetic field effect via charge-dependent directed flow in heavy-ion collisions at the Relativistic Heavy Ion Collider}",
    eprint = "2304.03430",
    archivePrefix = "arXiv",
    primaryClass = "nucl-ex",
    doi = "10.1103/PhysRevX.14.011028",
    journal = "Phys. Rev. X",
    volume = "14",
    number = "1",
    pages = "011028",
    year = "2024"
}

@article{Bozek:2022svy,
    author = "Bozek, Piotr",
    title = "{Splitting of proton-antiproton directed flow in relativistic heavy-ion collisions}",
    eprint = "2207.04927",
    archivePrefix = "arXiv",
    primaryClass = "nucl-th",
    doi = "10.1103/PhysRevC.106.L061901",
    journal = "Phys. Rev. C",
    volume = "106",
    number = "6",
    pages = "L061901",
    year = "2022"
}

@article{Guo:2012qi,
    author = "Guo, Yao and Liu, Feng and Tang, Aihong",
    title = "{Directed flow of transported and non-transported protons in Au+Au collisions from UrQMD model}",
    eprint = "1206.2246",
    archivePrefix = "arXiv",
    primaryClass = "nucl-ex",
    doi = "10.1103/PhysRevC.86.044901",
    journal = "Phys. Rev. C",
    volume = "86",
    pages = "044901",
    year = "2012"
}

@article{STAR:2017okv,
    author = "Adamczyk, Leszek and others",
    collaboration = "STAR",
    title = "{Beam-Energy Dependence of Directed Flow of $\Lambda$, $\bar{\Lambda}$, $K^\pm$, $K^0_s$ and $\phi$ in Au+Au Collisions}",
    eprint = "1708.07132",
    archivePrefix = "arXiv",
    primaryClass = "hep-ex",
    doi = "10.1103/PhysRevLett.120.062301",
    journal = "Phys. Rev. Lett.",
    volume = "120",
    number = "6",
    pages = "062301",
    year = "2018"
}

@article{STAR:2014clz,
    author = "Adamczyk, L. and others",
    collaboration = "STAR",
    title = "{Beam-Energy Dependence of the Directed Flow of Protons, Antiprotons, and Pions in Au+Au Collisions}",
    eprint = "1401.3043",
    archivePrefix = "arXiv",
    primaryClass = "nucl-ex",
    doi = "10.1103/PhysRevLett.112.162301",
    journal = "Phys. Rev. Lett.",
    volume = "112",
    number = "16",
    pages = "162301",
    year = "2014"
}

@article{Bilandzic:2010jr,
    author = "Bilandzic, Ante and Snellings, Raimond and Voloshin, Sergei",
    title = "{Flow analysis with cumulants: Direct calculations}",
    eprint = "1010.0233",
    archivePrefix = "arXiv",
    primaryClass = "nucl-ex",
    doi = "10.1103/PhysRevC.83.044913",
    journal = "Phys. Rev. C",
    volume = "83",
    pages = "044913",
    year = "2011"
}

@article{Poskanzer:1998yz,
    author = "Poskanzer, Arthur M. and Voloshin, S. A.",
    title = "{Methods for analyzing anisotropic flow in relativistic nuclear collisions}",
    eprint = "nucl-ex/9805001",
    archivePrefix = "arXiv",
    doi = "10.1103/PhysRevC.58.1671",
    journal = "Phys. Rev. C",
    volume = "58",
    pages = "1671--1678",
    year = "1998"
}

@article{Voloshin:1994mz,
    author = "Voloshin, S. and Zhang, Y.",
    title = "{Flow study in relativistic nuclear collisions by Fourier expansion of Azimuthal particle distributions}",
    eprint = "hep-ph/9407282",
    archivePrefix = "arXiv",
    doi = "10.1007/s002880050141",
    journal = "Z. Phys. C",
    volume = "70",
    pages = "665--672",
    year = "1996"
}

@article{Voloshin:2004vk,
    author = "Voloshin, Sergei A.",
    title = "{Parity violation in hot QCD: How to detect it}",
    eprint = "hep-ph/0406311",
    archivePrefix = "arXiv",
    doi = "10.1103/PhysRevC.70.057901",
    journal = "Phys. Rev. C",
    volume = "70",
    pages = "057901",
    year = "2004"
}

@article{Fukushima:2008xe,
    author = "Fukushima, Kenji and Kharzeev, Dmitri E. and Warringa, Harmen J.",
    title = "{The Chiral Magnetic Effect}",
    eprint = "0808.3382",
    archivePrefix = "arXiv",
    primaryClass = "hep-ph",
    doi = "10.1103/PhysRevD.78.074033",
    journal = "Phys. Rev. D",
    volume = "78",
    pages = "074033",
    year = "2008"
}

@article{Kharzeev:2010gd,
    author = "Kharzeev, Dmitri E. and Yee, Ho-Ung",
    title = "{Chiral Magnetic Wave}",
    eprint = "1012.6026",
    archivePrefix = "arXiv",
    primaryClass = "hep-th",
    reportNumber = "BNL-94527-2010-JA",
    doi = "10.1103/PhysRevD.83.085007",
    journal = "Phys. Rev. D",
    volume = "83",
    pages = "085007",
    year = "2011"
}

@article{Kharzeev:2007jp,
    author = "Kharzeev, Dmitri E. and McLerran, Larry D. and Warringa, Harmen J.",
    title = "{The Effects of topological charge change in heavy ion collisions: 'Event by event P and CP violation'}",
    eprint = "0711.0950",
    archivePrefix = "arXiv",
    primaryClass = "hep-ph",
    doi = "10.1016/j.nuclphysa.2008.02.298",
    journal = "Nucl. Phys. A",
    volume = "803",
    pages = "227--253",
    year = "2008"
}

@article{Voronyuk:2011jd,
    author = "Voronyuk, V. and Toneev, V. D. and Cassing, W. and Bratkovskaya, E. L. and Konchakovski, V. P. and Voloshin, S. A.",
    title = "{(Electro-)Magnetic field evolution in relativistic heavy-ion collisions}",
    eprint = "1103.4239",
    archivePrefix = "arXiv",
    primaryClass = "nucl-th",
    doi = "10.1103/PhysRevC.83.054911",
    journal = "Phys. Rev. C",
    volume = "83",
    pages = "054911",
    year = "2011"
}

@article{Tuchin:2010vs,
    author = "Tuchin, Kirill",
    title = "{Synchrotron radiation by fast fermions in heavy-ion collisions}",
    eprint = "1006.3051",
    archivePrefix = "arXiv",
    primaryClass = "nucl-th",
    reportNumber = "RBRC-845",
    doi = "10.1103/PhysRevC.83.039903",
    journal = "Phys. Rev. C",
    volume = "82",
    pages = "034904",
    year = "2010",
    note = "[Erratum: Phys.Rev.C 83, 039903 (2011)]"
}

@article{McLerran:2013hla,
    author = "McLerran, L. and Skokov, V.",
    title = "{Comments About the Electromagnetic Field in Heavy-Ion Collisions}",
    eprint = "1305.0774",
    archivePrefix = "arXiv",
    primaryClass = "hep-ph",
    reportNumber = "BNL-100762-2013-JA",
    doi = "10.1016/j.nuclphysa.2014.05.008",
    journal = "Nucl. Phys. A",
    volume = "929",
    pages = "184--190",
    year = "2014"
}

@article{Tuchin:2013ie,
    author = "Tuchin, Kirill",
    title = "{Particle production in strong electromagnetic fields in relativistic heavy-ion collisions}",
    eprint = "1301.0099",
    archivePrefix = "arXiv",
    primaryClass = "hep-ph",
    doi = "10.1155/2013/490495",
    journal = "Adv. High Energy Phys.",
    volume = "2013",
    pages = "490495",
    year = "2013"
}

@article{Deng:2012pc,
    author = "Deng, Wei-Tian and Huang, Xu-Guang",
    title = "{Event-by-event generation of electromagnetic fields in heavy-ion collisions}",
    eprint = "1201.5108",
    archivePrefix = "arXiv",
    primaryClass = "nucl-th",
    doi = "10.1103/PhysRevC.85.044907",
    journal = "Phys. Rev. C",
    volume = "85",
    pages = "044907",
    year = "2012"
}

@article{Skokov:2009qp,
    author = "Skokov, V. and Illarionov, A. Yu. and Toneev, V.",
    title = "{Estimate of the magnetic field strength in heavy-ion collisions}",
    eprint = "0907.1396",
    archivePrefix = "arXiv",
    primaryClass = "nucl-th",
    doi = "10.1142/S0217751X09047570",
    journal = "Int. J. Mod. Phys. A",
    volume = "24",
    pages = "5925--5932",
    year = "2009"
}

@article{ALICE:2013yba,
    author = "Abbas, E. and others",
    collaboration = "ALICE",
    title = "{Mid-rapidity anti-baryon to baryon ratios in pp collisions at $\sqrt{s}$ = 0.9, 2.76 and 7 TeV measured by ALICE}",
    eprint = "1305.1562",
    archivePrefix = "arXiv",
    primaryClass = "nucl-ex",
    reportNumber = "CERN-PH-EP-2013-080",
    doi = "10.1140/epjc/s10052-013-2496-5",
    journal = "Eur. Phys. J. C",
    volume = "73",
    pages = "2496",
    year = "2013"
}

@article{KHARZEEV2008227,
title = {The effects of topological charge change in heavy ion collisions: “Event by event P and CP violation”},
journal = {Nuclear Physics A},
volume = {803},
number = {3},
pages = {227-253},
year = {2008},
issn = {0375-9474},
doi = {https://doi.org/10.1016/j.nuclphysa.2008.02.298},
url = {https://www.sciencedirect.com/science/article/pii/S037594740800078X},
author = {Dmitri E. Kharzeev and Larry D. McLerran and Harmen J. Warringa}
}

@article{KHARZEEV1996238,
title = {Can gluons trace baryon number?},
journal = {Physics Letters B},
volume = {378},
number = {1},
pages = {238-246},
year = {1996},
issn = {0370-2693},
doi = {https://doi.org/10.1016/0370-2693(96)00435-2},
url = {https://www.sciencedirect.com/science/article/pii/0370269396004352},
author = {D. Kharzeev}
}

@article{Frenklakh:2024mgu,
    author = "Frenklakh, David and Kharzeev, Dmitri and Rossi, Giancarlo and Veneziano, Gabriele",
    title = "{Baryon-number {\textemdash} flavor separation in the topological expansion of QCD}",
    eprint = "2405.04569",
    archivePrefix = "arXiv",
    primaryClass = "hep-ph",
    reportNumber = "CERN-TH-2024-055",
    doi = "10.1007/JHEP07(2024)262",
    journal = "JHEP",
    volume = "07",
    pages = "262",
    year = "2024"
}

@article{PhysRevD.10.186,
  title = {Single-particle distribution in the hydrodynamic and statistical thermodynamic models of multiparticle production},
  author = {Cooper, Fred and Frye, Graham},
  journal = {Phys. Rev. D},
  volume = {10},
  issue = {1},
  pages = {186--189},
  numpages = {0},
  year = {1974},
  month = {Jul},
  publisher = {American Physical Society},
  doi = {10.1103/PhysRevD.10.186},
  url = {https://link.aps.org/doi/10.1103/PhysRevD.10.186}
}

@article{Bazavov:2017dus,
    author = "Bazavov, A. and others",
    title = "{The QCD Equation of State to $\mathcal{O}(\mu_B^6)$ from Lattice QCD}",
    eprint = "1701.04325",
    archivePrefix = "arXiv",
    primaryClass = "hep-lat",
    doi = "10.1103/PhysRevD.95.054504",
    journal = "Phys. Rev. D",
    volume = "95",
    number = "5",
    pages = "054504",
    year = "2017"
}

@article{Nayak:2019vtn,
    author = "Nayak, Kishora and Shi, Shusu and Xu, Nu and Lin, Zi-Wei",
    title = "{Energy dependence study of directed flow in Au+Au collisions using an improved coalescence in a multiphase transport model}",
    eprint = "1904.03863",
    archivePrefix = "arXiv",
    primaryClass = "nucl-ex",
    doi = "10.1103/PhysRevC.100.054903",
    journal = "Phys. Rev. C",
    volume = "100",
    number = "5",
    pages = "054903",
    year = "2019"
}

@article{Kharzeev:2000ph,
    author = "Kharzeev, Dmitri and Nardi, Marzia",
    title = "{Hadron production in nuclear collisions at RHIC and high density QCD}",
    eprint = "nucl-th/0012025",
    archivePrefix = "arXiv",
    doi = "10.1016/S0370-2693(01)00457-9",
    journal = "Phys. Lett. B",
    volume = "507",
    pages = "121--128",
    year = "2001"
}

@article{Gursoy:2020jso,
    author = {G{\"u}rsoy, U. and Kharzeev, D. E. and Marcus, E. and Rajagopal, K. and Shen, C.},
    editor = "Liu, Feng and Wang, Enke and Wang, Xin-Nian and Xu, Nu and Zhang, Ben-Wei",
    title = "{Charge-dependent flow induced by electromagnetic fields in heavy ion collisions}",
    eprint = "2002.12818",
    archivePrefix = "arXiv",
    primaryClass = "hep-ph",
    doi = "10.1016/j.nuclphysa.2020.121837",
    journal = "Nucl. Phys. A",
    volume = "1005",
    pages = "121837",
    year = "2021"
}

@article{Andronic:2008gu,
    author = "Andronic, A. and Braun-Munzinger, P. and Stachel, J.",
    title = "{Thermal hadron production in relativistic nuclear collisions: The Hadron mass spectrum, the horn, and the QCD phase transition}",
    eprint = "0812.1186",
    archivePrefix = "arXiv",
    primaryClass = "nucl-th",
    doi = "10.1016/j.physletb.2009.06.021",
    journal = "Phys. Lett. B",
    volume = "673",
    pages = "142--145",
    year = "2009",
    note = "[Erratum: Phys.Lett.B 678, 516 (2009)]"
}

@article{Gursoy:2014aka,
    author = "Gursoy, Umut and Kharzeev, Dmitri and Rajagopal, Krishna",
    title = "{Magnetohydrodynamics, charged currents and directed flow in heavy ion collisions}",
    eprint = "1401.3805",
    archivePrefix = "arXiv",
    primaryClass = "hep-ph",
    doi = "10.1103/PhysRevC.89.054905",
    journal = "Phys. Rev. C",
    volume = "89",
    number = "5",
    pages = "054905",
    year = "2014"
}

@article{Dubla:2020bdz,
    author = {Dubla, Andrea and G\"ursoy, Umut and Snellings, Raimond},
    title = "{Charge-dependent flow as evidence of strong electromagnetic fields in heavy-ion collisions}",
    eprint = "2009.09727",
    archivePrefix = "arXiv",
    primaryClass = "hep-ph",
    doi = "10.1142/S0217732320503241",
    journal = "Mod. Phys. Lett. A",
    volume = "35",
    number = "39",
    pages = "2050324",
    year = "2020"
}

@article{Broniowski:2001ei,
    author = "Broniowski, Wojciech and Florkowski, Wojciech",
    title = "{Geometric relation between centrality and the impact parameter in relativistic heavy ion collisions}",
    eprint = "nucl-th/0110020",
    archivePrefix = "arXiv",
    doi = "10.1103/PhysRevC.65.024905",
    journal = "Phys. Rev. C",
    volume = "65",
    pages = "024905",
    year = "2002"
}

\end{document}